\documentclass[11pt,a4paper]{article}
\usepackage{amsmath,amsthm,amsfonts,amssymb,mathrsfs}
\usepackage{graphicx}
\usepackage{subfigure}
\usepackage{tabularx}
\usepackage[latin1]{inputenc}
\usepackage[swedish,english]{babel} % engelska huvudspråk
\newcommand{\myvec}[1]{{\boldsymbol#1}}
\newcommand{\diag}{\text{{\rm diag}}\,}
\newcommand{\curl}{{\text{{\rm curl}}}}

\mathchardef\semic="303B
\newcommand{\sett}[2]{ \{ #1 \, \semic \, #2 \} }
\newcommand{\R}{{\mathbb R}}
\newcommand{\C}{{\mathbb C}}
\newcommand{\A}{{\mathbb A}}
\newcommand{\mH}{{\mathcal H}}
\newcommand{\dirac}{{\mathbf D}}

\begin{document}

\title{Comparison of integral equations for the Maxwell
       transmission problem with general permittivities}
\author{
  Johan Helsing\thanks{Centre for Mathematical Sciences, Lund
    University, Box 118, 221 00 Lund, Sweden (johan.helsing@math.lth.se).},~
  Anders Karlsson\thanks{Electrical and Information Technology, Lund
    University, Box 118, 221 00 Lund, Sweden 
    (anders.karlsson@eit.lth.se).},~~and
  Andreas Ros\'en\thanks{Mathematical Sciences, Chalmers University of
    Technology and University of Gothenburg, 412 96 G{\"o}teborg,
    Sweden (andreas.rosen@chalmers.se).}  }
\date{} 
\maketitle

\begin{abstract}
  Two recently derived integral equations for the Maxwell transmission
  problem are compared through numerical tests on simply connected
  axially symmetric domains for non-magnetic materials. The winning
  integral equation turns out to be entirely free from false
  eigenwavenumbers for any passive materials, also for purely negative
  permittivity ratios and in the static limit, as well as free from
  false essential spectrum on non-smooth surfaces. It also appears to
  be numerically competitive to all other available integral equation
  reformulations of the Maxwell transmission problem, despite using
  eight scalar surface densities.
\end{abstract}

\section{Introduction}

The Maxwell transmission problem is about determining the fields that
result when an incident time-harmonic electromagnetic wave is
scattered from and transmitted into a bounded dielectric object. Two
pioneering integral equation reformulations (IERs) of this problem,
which are still popular, are the Müller IER \cite[Section
23]{Muller69} and the Poggio-Miller-Chang-Harrington-Wu-Tsai (PMCHWT)
IER \cite{MillPogg78,WuTsai77}. These IERs are of Fredholm's second
and first kind, respectively. Over the years, much effort has been
made to find more efficient IERs. Focus has been on avoiding
dense-mesh and topological low-frequency breakdown, on avoiding false
resonances, and on providing unique solutions for wider ranges of
material parameters. Among later contributions we mention
\cite{EpsGreNei19,GaHaJeVo20,GaneshHawkinsVolkov19,HelsKarl20,HelsRose20,LaiOneil19,LiFuShank18,VicGreFer18}.

In this work we compare the numerical performance of two recently
derived IERs of the Maxwell transmission problem. The two IERs, which
are of Fredholm's second kind with singular integral operators, are
referred to as ``Dirac'' and ``HK 8-dens''. ``Dirac'' is derived
in~\cite{HelsRose20} by embedding the time-harmonic Maxwell's
equations into a Dirac equation and by tuning the six free parameters
of this equation as to optimize numerical performance. ``HK 8-dens''
is derived in~\cite{HelsKarl20} by extending a classic IER of the
Helmholtz transmission problem~\cite{KleiMart88} via the use of four
uniqueness parameters. Both our IERs use eight unknown scalar surface
densities for modeling. This is more than other popular IERs use.
Typical numbers are
four~\cite{EpsGreNei19,LaiOneil19,MillPogg78,Muller69,WuTsai77} or
six~\cite{GaHaJeVo20,LiFuShank18,VicGreFer18}. The major advantage
with our new IERs, however, is that they offer unique solutions, that
is they are free from false eigenwavenumbers, in particular for
plasmonic scattering. Another advantage, from a programming point of
view, is that both IERs require only bounded integral operators with
double and single layer type kernels, and in particular do not use
(compact differences of) hypersingular integral operators.

Uniqueness for a wide range of parameters is an important property of
an IER of a parameter-dependent partial differential equation (PDE)
for many reasons. First, one may actually be interested in solving the
PDE for a wide range of parameters. Then uniqueness is obviously
important. Second, even if one is not interested in a wide range of
parameters, non-uniqueness outside the parameter regime of interest
can seriously affect the conditioning of an IER inside the parameter
regime of interest. Third, theoretical studies of the solvability of a
PDE, for example the time-harmonic Maxwell's equations in Lipschitz
domains, are often based on IERs. Then it is crucial that the IER has
the same solvability properties as the PDE it models.

While ``Dirac'' and ``HK 8-dens'' are similar in many respects, there
are also differences. ``Dirac'' is derived in \cite{HelsRose20} for
general magnetic materials and assuming only Lipschitz regularity of
the surface of the scattering object. ``HK 8-dens'' has the advantage
that its surface densities have immediate interpretations in terms of
boundary limits of physical fields. Two of these surface densities are
always zero and are only needed, for uniqueness, in the main linear
system that is to be solved. They are omitted in field evaluations,
whose cost then are the same as for six-density IERs.

The original papers~\cite{HelsKarl20,HelsRose20} use mutually
different notation and contain numerical examples for reduced and two
dimensional versions of the IERs. The purpose of this work is to
present ``Dirac'' and ``HK 8-dens'' in a unified and
programming-friendly notation and to conduct a series of numerical
experiments for their full versions in three dimensions as to see
which IER is the most efficient. For this comparison, we limit the
discussion to non-magnetic materials, for which ``HK 8-dens'' is
formulated. Moreover, the free parameters in ``Dirac'' are likely to
need further tuning in the magnetic case, which will appear in a
forthcoming publication.

We conclude the paper in Section~\ref{sec:conclusion} by comparing
some salient properties of our IERs to those of two available
competitive IERs for the Maxwell transmission problem: the ``Debye''
and the ``DFIE'' IERs~\cite{EpsGreNei13,VicGreFer18}. With the aim of
achieving a comparison as accurate as possible, given the available
information, we base the discussion in Section~\ref{sec:conclusion} on
the following aspects of an IER. (A) Has it been used for Lipschitz
regular/piecewise smooth surfaces $\Gamma$? (B) What types of
operators does it employ? (C) How fast does it compute all the
scattered and transmitted fields? (D) How does it behave in the
quasi-static limit? (E) How stable is it? Does it have false
eigenwavenumbers, false near-eigenwavenumbers, or false essential
spectrum?

\section{Problem formulation}
\label{sec:probform}

\begin{figure}[t]
\centering
\includegraphics[height=50mm]{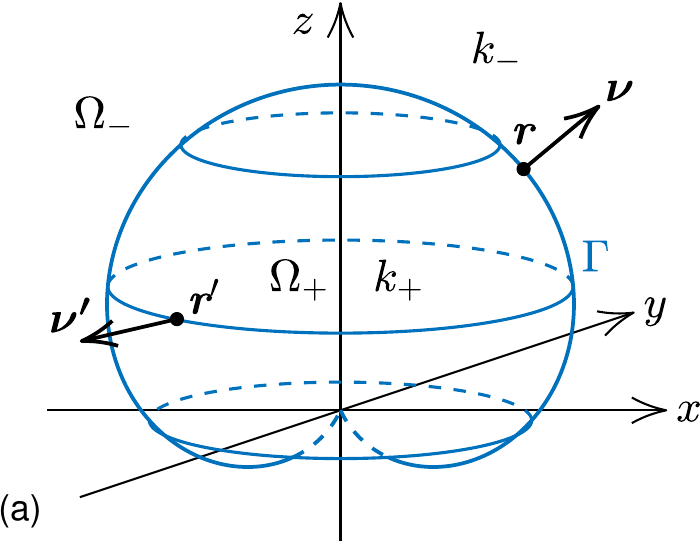}
  \hspace*{2mm}
\includegraphics[height=50mm]{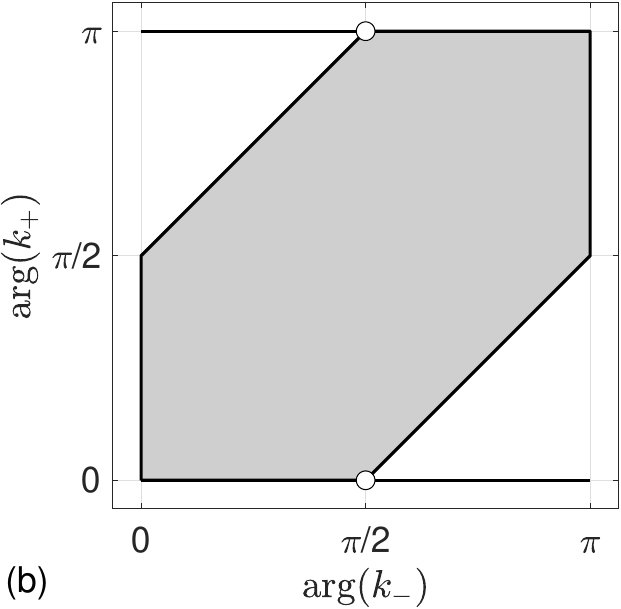}
\caption{\sf (a) Geometry in $\R^3$. Outside $\Gamma$ the volume is 
  $\Omega_-$ and the wavenumber $k_-$. Inside $\Gamma$ the volume is
  $\Omega_+$ and the wavenumber $k_+$. The outward unit normal is
  $\myvec \nu$ at $\myvec r$ and $\myvec \nu'$ at $\myvec r'$. (b) The
  gray region and the solid black lines constitute a set of points
  $(\arg(k_-),\arg(k_+))$ for which the Maxwell transmission problem
  has at most one solution. Circles are not included.}
\label{fig:amoeba0}
\end{figure}

Let $\Omega_+$ be a bounded domain in $\R^3$ with boundary surface
$\Gamma$ and an unbounded, connected, exterior $\Omega_-$. The outward
unit normal at position $\myvec r$ on $\Gamma$ is $\myvec\nu$. We
consider time-harmonic fields with time dependence $e^{-{\rm i}\omega
  t}$ and angular frequency $\omega>0$. The relation between
time-dependent fields $F(\myvec r,t)$ and complex fields $F(\myvec r)$
is
\begin{equation} 
F(\myvec r,t)=\Re{\rm e}\left\{F(\myvec r)e^{-{\rm i}\omega t}\right\}.
\label{eq:timedep}
\end{equation}
The domains $\Omega_\pm$ are homogeneous with material properties
described by wavenumbers $k_\pm$. See Figure~\ref{fig:amoeba0}(a).
The $k_\pm$ are related to the total permittivities $\epsilon_\pm$ and
permeabilities $\mu_\pm$ by $k_\pm=\omega\sqrt{\epsilon_\pm\mu_\pm}$.
To avoid issues regarding the choice of branch of the square root, the
$k_\pm$ are considered our basic parameters. Passive materials have
$\Im{\rm m}\{\epsilon_\pm\}\ge 0$ and $\Im{\rm m}\{\mu_\pm\}\ge 0$ and
values of $k_\pm$ in the closed first quadrant. In Maxwell's
equations, $\nabla\times {\mathcal E}= {\rm i}\omega\mu_\pm {\mathcal
  H}$ and $\nabla\times {\mathcal H}= -{\rm i}\omega\epsilon_\pm
{\mathcal E}$ in $\Omega_\pm$, we rescale the magnetic field
${\mathcal H}$ by the wave impedance in $\Omega_-$ and set $\myvec
E={\mathcal E}$ and $\myvec H= \sqrt{\mu_-/\epsilon_-} {\mathcal H}$
in all $\R^3$. Throughout the paper, only non-magnetic materials are
considered. That is, $\mu_\pm=\mu_0$, where $\mu_0$ is the
permeability of vacuum. We now state our formulation of the Maxwell
transmission problem:

Given incident fields $\myvec E^{\rm in}$ and $\myvec H^{\rm in}$,
generated in $\Omega_-$, we seek the total electric and magnetic
fields $\myvec E(\myvec r)$ and $\myvec H(\myvec r)$, $\myvec
r\in\Omega_-\cup\Omega_+$, which, for $(k_-,k_+)$ with arguments in
the set of points shown in Figure~\ref{fig:amoeba0}(b) and with
$\hat{\epsilon}$ such that
\begin{equation}
\hat{\epsilon}\equiv\epsilon_+/\epsilon_-={k_+^2}/{k_-^2}
\quad\mbox{and}\quad
\hat{\epsilon}\neq -1\,,
\label{eq:kappa}
\end{equation}
solve Maxwell's equations
\begin{equation}
\begin{split}
\nabla\times\myvec E(\myvec r)&= {\rm i}k_-\myvec H(\myvec r)\,,
\quad\myvec r\in\Omega_-\cup\Omega_+\,,\\
\nabla\times\myvec H(\myvec r)&=-{\rm i}k_-\myvec E(\myvec r)\,,
\quad\myvec r\in\Omega_-\,,\\
\nabla\times\myvec H(\myvec r)&=-{\rm i}k_-\hat{\epsilon}\myvec E(\myvec r)\,,
\quad\myvec r\in\Omega_+\,,
\end{split}
\label{eq:Max123C}
\end{equation}
except possibly at an isolated point in $\Omega_-$ where the source of
$\myvec E^{\rm in}$ and $\myvec H^{\rm in}$ is located, subject to the
boundary conditions
\begin{align}
\lim_{\Omega_-\ni\myvec r'\to\myvec r}
\myvec\nu\times\myvec E(\myvec r')&=
\lim_{\Omega_+\ni\myvec r'\to\myvec r}
\myvec\nu\times\myvec E(\myvec r')\,,
\quad\myvec r\in\Gamma\,,\label{eq:rv2C}\\
\lim_{\Omega_-\ni\myvec r'\to\myvec r}
\myvec\nu\times\myvec H(\myvec r')&=
\lim_{\Omega_+\ni\myvec r'\to\myvec r}
\myvec\nu\times\myvec H(\myvec r')\,,
\quad\myvec r\in\Gamma\,,\label{eq:rv1C}\\
{\myvec r/|\myvec r|}\times\myvec E^{\rm sc}(\myvec r)-\myvec H^{\rm sc} 
(\myvec r)&=o\left(\vert\myvec r\vert^{-1}e^{\Im\rm{m}\{k_-\} 
\vert\myvec r\vert}\right),\quad\lvert\myvec r\rvert\rightarrow\infty\,,\\
{\myvec r/|\myvec r|}\times\myvec H^{\rm sc}(\myvec r)+\myvec H^{\rm sc} 
(\myvec r)&=o\left(\vert\myvec r\vert^{-1}e^{\Im\rm{m}\{k_-\}
\vert\myvec r\vert}\right),\quad\lvert\myvec r\rvert\rightarrow\infty\,.
\label{eq:radcondC}
\end{align}
The scattered fields $\myvec E^{\rm sc}$ and $\myvec H^{\rm sc}$ are
source free in $\Omega_-$ and defined, along with the transmitted
fields $\myvec E^{\rm tr}$ and $\myvec H^{\rm tr}$, by
\begin{equation}
\begin{split}
\myvec E(\myvec r)&=\left\{
\begin{array}{ll}
\myvec E^{\rm in}(\myvec r)+\myvec E^{\rm sc}(\myvec r)\,, 
& \myvec r\in\Omega_-\,,\\
\myvec E^{\rm tr}(\myvec r)\,, & \myvec r\in\Omega_+\,.
\end{array}
\right.\\
\myvec H(\myvec r)&=\left\{
\begin{array}{ll}
\myvec H^{\rm in}(\myvec r)+\myvec H^{\rm sc}(\myvec r)\,, 
& \myvec r\in\Omega_-\,,\\
\myvec H^{\rm tr}(\myvec r)\,, & \myvec r\in\Omega_+\,.
\end{array}
\right.
\end{split}
\label{eq:decompEH}
\end{equation}
The incident fields satisfy
\begin{equation}
\begin{split}
\nabla\times\myvec E^{\rm in}(\myvec r)&= {\rm i}k_-\myvec H^{\rm in}
(\myvec r)\,,\quad\myvec r\in\R^3\,,\\
\nabla\times\myvec H^{\rm in}(\myvec r)&=-{\rm i}k_-\myvec E^{\rm in}
(\myvec r)\,,\quad\myvec r\in\R^3\,,
\end{split}
\end{equation}
except at the possible isolated source point in $\Omega_-$. In
addition, conservation of charge must hold and by that
\begin{equation}
\int_\Gamma\myvec\nu\cdot\myvec E^{{\rm sc}-}(\myvec r)\,{\rm d}\Gamma=0\,,
\label{eq:charge}
\end{equation}
where $\myvec E^{{\rm sc}-}$ is the exterior limit of $\myvec E^{\rm
  sc}$ on $\Gamma$. In what follows, the Maxwell transmission
problem~(\ref{eq:Max123C})--(\ref{eq:charge}) will be referred to as
{\em the MTP($k_-,k_+$)}.

\section{Properties of IERs of the MTP($k_-,k_+$)}
\label{sec:prop}

Let us first mention that there is yet no IER of the MTP($k_-,k_+$)
known that is equivalent to the MTP($k_-,k_+$) itself for all pairs of
complex-valued $(k_-,k_+)$, not even for all
$0\leq\arg(k_-),\arg(k_+)\leq\pi$. Neither is it known for which
$(k_-,k_+)$ there exist unique solutions to the MTP($k_-,k_+$). The
most comprehensive result we are aware of, for uniqueness with
Lipschitz regular $\Gamma$ and with $\mu=1$, says that the
MTP($k_-,k_+$) has at most one solution when $(\arg(k_-),\arg(k_+))$
belongs to the set of points shown in
Figure~\ref{fig:amoeba0}(b)~\cite[Proposition 8.2]{HelsRose20}. As for
existence of solutions, the same set of points apply with the
restriction that $\hat{\epsilon}\ne -1$ for smooth $\Gamma$ and that
$\hat{\epsilon}$ is outside of a geometry-dependent interval on the
negative real axis, containing $\hat{\epsilon}=-1$, for merely
Lipschitz regular $\Gamma$~\cite[Proposition 8.3]{HelsRose20}.

For ``Dirac'' it is proven that there exist unique solutions when
$\Omega_+$ is a bounded Lipschitz domain with connected exterior
$\Omega_-$ and $(k_-,k_+)$ satisfies the conditions
of~\cite[Proposition 8.3]{HelsRose20}. Existence of unique solutions
for ``HK 8-dens" is proven for $(k_-,k_+)$ as described in
\cite[Section 8.3]{HelsKarl20}. It is assumed in \cite{HelsKarl20}
that $\Gamma$ is smooth and $\Omega_+$ simply connected but the proof
holds also for objects that are multiply connected. All other IERs of
the MTP($k_-,k_+$) in the electromagnetics literature, possibly with
the exception of the Debye representation~\cite{EpsGreNei13}, seem to
have unique solutions only under more restrictive conditions. In
particular, they can not guarantee uniqueness for any
$(\arg(k_-),\arg(k_+))=(0,\pi/2)$.

Two problems shared by many IERs of the MTP($k_-,k_+$), which have
received much attention recently, are {\it dense-mesh and topological
  low-frequency breakdown}, see~\cite{VicGreFer18} and~\cite[Section
2]{EpsGreNei19}. Dense-mesh low-frequency breakdown refers to
catastrophic cancellation that destroys the numerical accuracy in the
computed fields in the static limit. Topological low-frequency
breakdown is a, perhaps, more elusive phenomenon that broadly seems to
refer to an increased ill-conditioning of the integral equation itself
in the quasi-static limit $k_\pm\to 0$, with $\hat
\epsilon=(k_+/k_-)^2$ fixed, for $\Gamma$ with non-zero genus. For
scattering by superconductors, this phenomenon is discussed in
\cite{AndiulliETAL09,EpsteinETAL13}. ``Dirac'' is proven to be free
from dense-mesh low-frequency breakdown and, at large, also from
topological low-frequency breakdown. See Section~\ref{sec:static}
below.

Additional problems, which have not received as much attention as
low-frequency breakdown but still cause numerical degradation of IERs,
include {\em false near-eigenwavenumbers} and {\em false essential
  spectrum}. An {\em eigenwavenumber} is a pair of wavenumbers
$(k_-,k_+)$ for which the IER does not have a unique solution. If the
MTP($k_-,k_+$) has a unique solution we speak of a false
eigenwavenumber, whereas we call it a true eigenwavenumber if even the
MTP($k_-,k_+$) fails to have a unique solution. If we only consider
wavenumber pairs from a set $X$ and there is an eigenwavenumber $z$
outside but close to $X$, then a pair $x\in X$ near $z$ which locally
maximizes the condition number of the IER, is referred to a {\em
  near-eigenwavenumber}. This can be a true or false such, depending
on the nature of $z$. The terminology of {\em true and false
  eigenwavenumbers} was introduced in~\cite[Section 5.3]{HelsKarl18},
to make more precise the common terminology of {\em true and spurious
  resonances}. The term {\em spurious near resonances} is used in
\cite{VicGreFer18}.

False essential spectrum may appear for certain $(k_-^{\rm f},k_+^{\rm
  f})$ with $\hat{\epsilon}^{\rm f}=(k_+^{\rm f}/k_-^{\rm f})^2$ real
and negative and a merely Lipschitz regular $\Gamma$. More precisely:
for a pair of wavenumbers $(k_-^{\rm f},k_+^{\rm f})$, we say that the
IER has false essential spectrum if it is not a Fredholm operator,
even though the MTP($k_-^{\rm f},k_+^{\rm f}$) defines a Fredholm map.
While the term {\em essential spectrum} is standard, the term {\em
  false/spurious essential spectrum} does not seem to have been used.

Let $k_-=k_-^{\rm f}$ and $k_+\to k_+^{\rm f}$ in such a way that
$\hat{\epsilon}\to\hat{\epsilon}^{\rm f}$ from above or from below in
the complex plane. At a point $(k_-^{\rm f},k_+^{\rm f})$ where we
have false essential spectrum, the typical numerical behavior of the
IER is that it has the same unique limit solution from above and below
and this solution coincides with the solution to the MTP($k_-^{\rm
  f},k_+^{\rm f}$), while the IER is not solvable for
$(k_-,k_+)=(k_-^{\rm f},k_+^{\rm f})$.

\section{Notation for axially symmetric $\Gamma$}
\label{sec:coordvec}

\begin{figure}[t!]
\centering
\includegraphics[height=44mm]{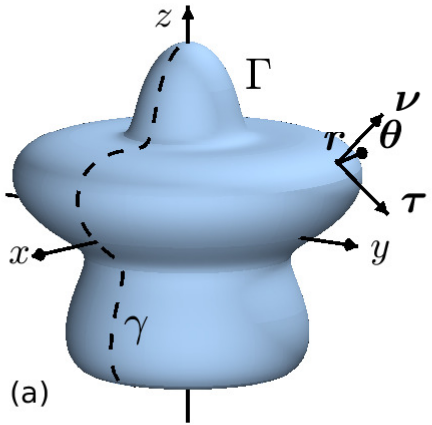}
  \hspace*{2mm}
\includegraphics[height=44mm]{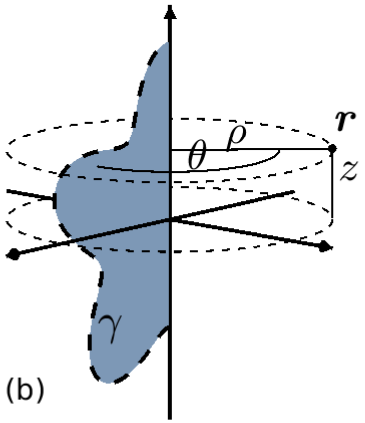}
  \hspace*{2mm}
\includegraphics[height=44mm]{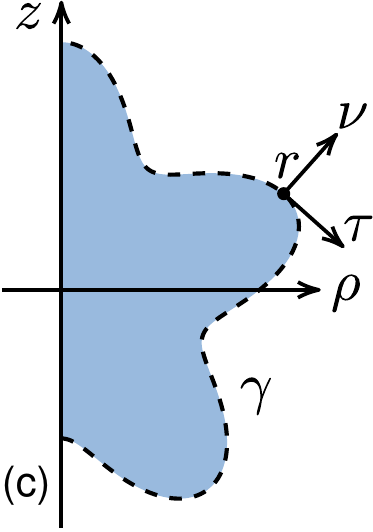}
\caption{\sf Notation for an axially symmetric surface $\Gamma$
  generated by a curve $\gamma$: (a) a point $\myvec r$ on $\Gamma$
  has outward unit normal $\myvec\nu$ and unit tangent vectors
  $\myvec\tau$ and $\myvec\theta$; (b) cylindrical coordinates
  $(\rho,\theta,z)$ of $\myvec r$; (c) coordinate axes and vectors in
  the half-plane $\mathbb{A}$ defined by $\theta=0$.}
\label{fig:notation}
\end{figure}

From now on we assume that $\Gamma$ is axially symmetric, since our
numerical examples cover such domains. Note, however, that the full 3D
formulations of our IERs and of the representations of $\myvec E$ and
$\myvec H$ are given in \cite{HelsKarl20,HelsRose20}. These IERs are
in no way limited to axially symmetric domains. In the present paper
we use both Cartesian coordinates $x$, $y$, $z$ and cylindrical
coordinates $\rho$, $\theta$, $z$, as illustrated in
Figure~\ref{fig:notation}. A half-plane $\A$ in $\R^2$ is defined by
$\theta=0$, a generating curve $\gamma$ is defined by the intersection
of $\Gamma$ and $\A$, and a general point in $\A$ is
\begin{equation}
r=(\rho,z)\,.
\label{eq:2D1}
\end{equation}
The outward unit normal and a tangent on $\gamma$, in $\A$, are
\begin{align}
\nu&=(\nu_\rho,\nu_z)\,,
\label{eq:2D2}\\
\tau&=(\nu_z,-\nu_\rho)\,.
\label{eq:2D3}
\end{align}
In $\R^3$, the position with Cartesian basis vectors is
\begin{equation}
\myvec r\equiv (x,y,z)=(\rho\cos\theta,\rho\sin\theta,z)\,.
\end{equation}
We also use the cylindrical unit vectors
\begin{align}
\myvec\rho&=(\cos\theta,\sin\theta,0)\,,\\
\myvec\theta&=(-\sin\theta,\cos\theta,0)\,,\\
\myvec z&=(0,0,1)\,.
\end{align}
On $\Gamma$, the unit normal is $\myvec\nu$ and the tangential unit
vectors are $\myvec\tau$ and $\myvec\theta$
\begin{align}
\myvec\nu&=(\nu_\rho\cos\theta,\nu_\rho\sin\theta,\nu_z)\,,\\
\myvec\tau&\equiv\myvec\theta\times\myvec\nu=
(\nu_z\cos\theta,\nu_z\sin\theta,-\nu_\rho)\,.
\end{align}
Vector fields, off from $\Gamma$, will often be expressed using
$\myvec\rho$, $\myvec\theta$, and $\myvec z$
\begin{equation}
\begin{split}
\myvec E&=\myvec\rho E_{\rho}+\myvec\theta E_{\theta}+\myvec zE_z\,,\\
\myvec H&=\myvec\rho H_{\rho}+\myvec\theta H_{\theta}+\myvec zH_z\,.
\end{split}
\label{eq:EHcylind}
\end{equation}
The causal fundamental solution to the Helmholtz' equation and its
gradient, renormalized by a factor of $-2$ to make the expressions of
the integral operators in Appendix~\ref{app:A} simpler, are
\begin{align}
\Phi_k(\myvec r,\myvec r')&=\frac{e^{{\rm i}k\vert\myvec r-\myvec r'\vert}}
{2\pi\vert\myvec r-\myvec r'\vert}\,,
\label{eq:funda}\\
\nabla'\Phi_k(\myvec r,\myvec r')&=\frac{\myvec r-\myvec r'}
{2\pi\vert\myvec r-\myvec r'\vert^3}
(1-{\rm i}k\vert\myvec r-\myvec r'\vert)
e^{{\rm i}k\vert \myvec r-\myvec r'\vert}\,.
\label{eq:gradfunda}
\end{align}

\section{A unified formalism}

The integral equations of ``Dirac'' and ``HK 8-dens'' can both be
written in the general form
\begin{equation}
  (I+G)h(\myvec r)= 2Nf^{\rm in}(\myvec r)\,,\quad\myvec r\in\Gamma\,,
\label{eq:3Dgen}
\end{equation}
with
\begin{equation}
  G=PE_{k_+}N'-NE_{k_-}P'\,.
\label{eq:G}
\end{equation}
Here $h$ contains eight unknown scalar surface densities
\begin{equation}
h=\begin{bmatrix}
h_1 & h_2 & h_3 & h_4 & h_5 & h_6 & h_7 & h_8
\end{bmatrix}^{\rm T},
\label{eq:h}
\end{equation}
$f^{\rm in}$ contains the field components
\begin{equation}
f^{\rm in}=
\begin{bmatrix}
0 & \myvec\nu\cdot\myvec H^{\rm in}  & \myvec\tau\cdot\myvec H^{\rm in}
  & \myvec\theta\cdot\myvec H^{\rm in} &
0 & \myvec\nu\cdot\myvec E^{\rm in}  & \myvec\tau\cdot\myvec E^{\rm in}
  & \myvec\theta\cdot\myvec E^{\rm in}
\end{bmatrix}^{\rm T},
  \label{eq:maxwrhs}
\end{equation}
$E_k$ is the Cauchy singular $8\times 8$ block operator matrix
\begin{equation}
\begin{bmatrix}
-K^{\myvec\nu'}_k  & 0 & K^{\myvec\theta'}_k & -K^{\myvec\tau'}_k &
 0 & S^1_k & 0 & 0  \\
 K^{\myvec\nu\times\myvec\nu'}_k & -K^\myvec\nu_k & 
-K^{\myvec\nu\times\myvec\theta'}_k &
 K^{\myvec\nu\times\myvec\tau'}_k & S^{\myvec\nu\cdot\myvec\nu'}_k &
 0 & S^{\myvec\nu\cdot\myvec\theta'}_k & -S^{\myvec\nu\cdot \myvec\tau'}_k\\
 K^{\myvec\tau\times\myvec\nu'}_k & -K^\myvec\tau_k & 
-K^{\myvec\tau\times\myvec\theta'}_k & K^{\myvec\tau\times\myvec\tau'}_k & 
 S^{\myvec\tau\cdot\myvec\nu'}_k & 0 & S^{\myvec\tau\cdot\myvec\theta'}_k &
-S^{\myvec\tau\cdot\myvec\tau'}_k\\
 K^{\myvec\theta\times\myvec\nu'}_k & -K^\myvec\theta_k &
-K^{\myvec\theta\times\myvec\theta'}_k & 
 K^{\myvec\theta\times\myvec\tau'}_k & S^{\myvec\theta\cdot\myvec\nu'}_k & 
 0 & S^{\myvec\theta\cdot\myvec\theta'}_k & 
-S^{\myvec\theta\cdot\myvec\tau'}_k\\
 0 & S^1_k & 0 & 0 & -K^{\myvec\nu'}_k & 0 & -K^{\myvec\theta'}_k & 
 K^{\myvec\tau'}_k \\
 S^{\myvec\nu\cdot\myvec\nu'}_k & 0 & -S^{\myvec\nu\cdot\myvec\theta'}_k & 
 S^{\myvec\nu\cdot\myvec\tau'}_k & -K^{\myvec\nu\times\myvec\nu'}_k & 
-K^\myvec\nu_k & -K^{\myvec\nu\times\myvec\theta'}_k & 
 K^{\myvec\nu\times\myvec\tau'}_k\\
 S^{\myvec\tau\cdot\myvec\nu'}_k & 0 & -S^{\myvec\tau\cdot\myvec\theta'}_k & 
 S^{\myvec\tau\cdot\myvec\tau'}_k & -K^{\myvec\tau\times\myvec\nu'}_k & 
-K^\myvec\tau_k & -K^{\myvec\tau\times\myvec\theta'}_k & 
 K^{\myvec\tau\times\myvec\tau'}_k\\
 S^{\myvec\theta\cdot\myvec\nu'}_k & 0 & 
-S^{\myvec\theta\cdot\myvec\theta'}_k & 
 S^{\myvec\theta\cdot \myvec\tau'}_k & -K^{\myvec\theta\times\myvec\nu'}_k & 
-K^\myvec\theta_k & -K^{\myvec\theta\times\myvec\theta'}_k & 
 K^{\myvec\theta\times\myvec\tau'}_k  
\end{bmatrix}
\label{eq:Ek3D}
\end{equation}
with operator entries detailed in Appendix~\ref{app:A}, and $P$, $P'$,
$N$, $N'$ are diagonal matrices which are specified by 
``Dirac'' and ``HK 8-dens'', respectively.

Once~(\ref{eq:3Dgen}) is solved for $h$, 
the fields $\myvec E$ and $\myvec H$ in $\Omega_\pm$,
decomposed as
in~(\ref{eq:EHcylind}), can be evaluated from~(\ref{eq:decompEH}) and
\begin{equation}
\begin{split}
E_\rho^{\rm sc}&=\frac{1}{2}
\begin{bmatrix}
 \tilde S_{k_-}^{\myvec\rho\cdot\myvec\nu'}& 
 0&
-\tilde S_{k_-}^{\myvec\rho\cdot\myvec\theta'}&
 \tilde S_{k_-}^{\myvec\rho\cdot\myvec\tau'}&
-\tilde K_{k_-}^{\myvec\rho\times\myvec\nu'}&
-\tilde K_{k_-}^{\myvec\rho}&
-\tilde K_{k_-}^{\myvec\rho\times\myvec\theta'}&
 \tilde K_{k_-}^{\myvec\rho\times\myvec\tau'}
\end{bmatrix}h^-,\\
E_\theta^{\rm sc}&=\frac{1}{2}
\begin{bmatrix}
 \tilde S_{k_-}^{\myvec\theta\cdot\myvec\nu'}& 
 0&
-\tilde S_{k_-}^{\myvec\theta\cdot\myvec\theta'}&
 \tilde S_{k_-}^{\myvec\theta\cdot\myvec\tau'}&
-\tilde K_{k_-}^{\myvec\theta\times\myvec\nu'}&
-\tilde K_{k_-}^{\myvec\theta}&
-\tilde K_{k_-}^{\myvec\theta\times\myvec\theta'}&
 \tilde K_{k_-}^{\myvec\theta\times\myvec\tau'}
\end{bmatrix}h^-,\\
E_z^{\rm sc}&=\frac{1}{2}
\begin{bmatrix}
 \tilde S_{k_-}^{\myvec z\cdot\myvec\nu'}& 
 0&\quad\;\;\;0\;\;&
 \tilde S_{k_-}^{\myvec z\cdot\myvec\tau'}&
-\tilde K_{k_-}^{\myvec z\times\myvec\nu'}&
-\tilde K_{k_-}^{\myvec z}&
-\tilde K_{k_-}^{\myvec z\times\myvec\theta'}&
 \tilde K_{k_-}^{\myvec z\times\myvec\tau'}
\end{bmatrix}h^-,
\end{split}
\label{eq:Efieldminus}
\end{equation}
\begin{equation}
\begin{split}
E_\rho^{\rm tr}&=\frac{1}{2}
\begin{bmatrix}
 \tilde S_{k_+}^{\myvec\rho\cdot\myvec\nu'}& 
 0&
-\tilde S_{k_+}^{\myvec\rho\cdot\myvec\theta'}&
 \tilde S_{k_+}^{\myvec\rho\cdot\myvec\tau'}&
-\tilde K_{k_+}^{\myvec\rho\times\myvec\nu'}&
-\tilde K_{k_+}^{\myvec\rho}&
-\tilde K_{k_+}^{\myvec\rho\times\myvec\theta'}&
 \tilde K_{k_+}^{\myvec\rho\times\myvec\tau'}
\end{bmatrix}h^+,\\
E_\theta^{\rm tr}&=\frac{1}{2}
\begin{bmatrix}
 \tilde S_{k_+}^{\myvec\theta\cdot\myvec\nu'}& 
 0&
-\tilde S_{k_+}^{\myvec\theta\cdot\myvec\theta'}&
 \tilde S_{k_+}^{\myvec\theta\cdot\myvec\tau'}&
-\tilde K_{k_+}^{\myvec\theta\times\myvec\nu'}&
-\tilde K_{k_+}^{\myvec\theta}&
-\tilde K_{k_+}^{\myvec\theta\times\myvec\theta'}&
 \tilde K_{k_+}^{\myvec\theta\times\myvec\tau'}
\end{bmatrix}h^+,\\
E_z^{\rm tr}&=\frac{1}{2}
\begin{bmatrix}
 \tilde S_{k_+}^{\myvec z\cdot\myvec\nu'}& 
 0&\quad\;\;\;0\;\;&
 \tilde S_{k_+}^{\myvec z\cdot\myvec\tau'}&
-\tilde K_{k_+}^{\myvec z\times\myvec\nu'}&
-\tilde K_{k_+}^{\myvec z}&
-\tilde K_{k_+}^{\myvec z\times\myvec\theta'}&
 \tilde K_{k_+}^{\myvec z\times\myvec\tau'}
\end{bmatrix}h^+,
\end{split}
\label{eq:Efieldplus}
\end{equation}

\begin{equation}
\begin{split}
H_{\rho}^{\rm sc}&=\frac{1}{2}
\begin{bmatrix}
 \tilde K_{k_-}^{\myvec\rho\times\myvec\nu'}&
-\tilde K_{k_-}^{\myvec\rho}&
-\tilde K_{k_-}^{\myvec\rho\times\myvec\theta'}&
 \tilde K_{k_-}^{\myvec\rho\times\myvec\tau'}&
 \tilde S_{k_-}^{\myvec\rho\cdot\myvec\nu'}&
 0&
\tilde S_{k_-}^{\myvec\rho\cdot\myvec\theta'}&
-\tilde S_{k_-}^{\myvec\rho\cdot\myvec\tau'}
\end{bmatrix}h^-,\\
H_{\theta}^{\rm sc}&=\frac{1}{2}
\begin{bmatrix}
 \tilde K_{k_-}^{\myvec\theta\times\myvec\nu'}&
-\tilde K_{k_-}^{\myvec\theta}&
-\tilde K_{k_-}^{\myvec\theta\times\myvec\theta'}&
 \tilde K_{k_-}^{\myvec\theta\times\myvec\tau'}&
 \tilde S_{k_-}^{\myvec\theta\cdot\myvec\nu'}&
 0&
\tilde S_{k_-}^{\myvec\theta\cdot\myvec\theta'}&
-\tilde S_{k_-}^{\myvec\theta\cdot\myvec\tau'}
\end{bmatrix}h^-,\\
H_{z}^{\rm sc}&=\frac{1}{2}
\begin{bmatrix}
 \tilde K_{k_-}^{\myvec z\times\myvec\nu'}&
-\tilde K_{k_-}^{\myvec z}&
-\tilde K_{k_-}^{\myvec z\times\myvec\theta'}&
 \tilde K_{k_-}^{\myvec z\times\myvec\tau'}&
 \tilde S_{k_-}^{\myvec z\cdot\myvec\nu'}&
 0&\quad 0\;\;&
-\tilde S_{k_-}^{\myvec z\cdot\myvec\tau'}
\end{bmatrix}h^-,
\end{split}
\label{eq:Hfieldminus}
\end{equation}
\begin{equation}
\begin{split}
H_{\rho}^{\rm tr}&=\frac{\hat k}{2}
\begin{bmatrix}
 \tilde K_{k_+}^{\myvec\rho\times\myvec\nu'}&
-\tilde K_{k_+}^{\myvec\rho}&
-\tilde K_{k_+}^{\myvec\rho\times\myvec\theta'}&
 \tilde K_{k_+}^{\myvec\rho\times\myvec\tau'}&
 \tilde S_{k_+}^{\myvec\rho\cdot\myvec\nu'}&
 0&
 \tilde S_{k_+}^{\myvec\rho\cdot\myvec\theta'}&
-\tilde S_{k_+}^{\myvec\rho\cdot\myvec\tau'}
\end{bmatrix}h^+,\\
H_{\theta}^{\rm tr}&=\frac{\hat k}{2}
\begin{bmatrix}
 \tilde K_{k_+}^{\myvec\theta\times\myvec\nu'}&
-\tilde K_{k_+}^{\myvec\theta}&
-\tilde K_{k_+}^{\myvec\theta\times\myvec\theta'}&
 \tilde K_{k_+}^{\myvec\theta\times\myvec\tau'}&
 \tilde S_{k_+}^{\myvec\theta\cdot\myvec\nu'}&
 0&
 \tilde S_{k_+}^{\myvec\theta\cdot\myvec\theta'}&
-\tilde S_{k_+}^{\myvec\theta\cdot\myvec\tau'}
\end{bmatrix}h^+,\\
H_{z}^{\rm tr}&=\frac{\hat k}{2}
\begin{bmatrix}
 \tilde K_{k_+}^{\myvec z\times\myvec\nu'}&
-\tilde K_{k_+}^{\myvec z}&
-\tilde K_{k_+}^{\myvec z\times\myvec\theta'}&
 \tilde K_{k_+}^{\myvec z\times\myvec\tau'}&
 \tilde S_{k_+}^{\myvec z\cdot\myvec\nu'}&
 0&\quad 0\;\;&
-\tilde S_{k_+}^{\myvec z\cdot\myvec\tau'}
\end{bmatrix}h^+.
\end{split}
\label{eq:Hfieldplus}
\end{equation} 
Here the layer-potential entries are detailed in Appendix~\ref{app:B}
and 
\begin{gather}
  \hat{k}=k_+/k_-\,,\\
  h^+=N'h\,,\qquad h^-=P'h\,.
\label{eq:regdens}
\end{gather}
Note that the additional zero entry of $E_z$ and $H_z$ is due to axial
symmetry.

For ``Dirac'', the field evaluation can be improved by instead using
the projected densities
 \begin{equation} \label{eq:tweakeddens}
\qquad h^+=\tfrac 12(I+ E_{k_+}) N'h\,,\qquad h^-=\tfrac 12(I-E_{k_-})P'h
\end{equation}
as input to the field evaluation
\eqref{eq:Efieldminus}--\eqref{eq:Hfieldplus}. The two choices
\eqref{eq:regdens} and \eqref{eq:tweakeddens} will always produce the
same fields in $\Omega_\pm$, and we discuss their uses in the
following subsections.

We remark that the framework~\eqref{eq:3Dgen}--\eqref{eq:Ek3D} applies
to both ``Dirac'' and to ``HK 8-dens'', also for general object shapes
$\Gamma$, whenever $\myvec\tau$ and $\myvec\theta$ are two arbitrary
orthogonal tangential unit vectors. The corresponding representations
of $\myvec E$ and $\myvec H$ are then given by \cite[Eqs.~(22) and
(23)]{HelsRose20}.

\subsection{Choice of $P$, $P'$, $N$, $N'$ for ``Dirac''}
\label{sec:Dirac}

Given the formalism above, ``Dirac'' specifies the diagonal
matrices
\begin{equation}  
\begin{split}  
P &=\diag\begin{bmatrix} 
    \frac{1-{\rm i}\delta\arg(\hat c)}{{\hat c}+1-{\rm i}\delta\arg(\hat c)} 
  & \frac{1}{\sqrt{{\hat c}+|{\hat c}|}}
  & \frac{1}{2\sqrt{\hat c}} & \frac{1}{2\sqrt{\hat c}} 
  & \frac{|{\hat c}|}{{\hat c}+|{\hat c}|} 
  & \frac{\hat\epsilon}{\hat\epsilon+1} & 1 & 1 
\end{bmatrix},  \\
P'&=\diag\begin{bmatrix} 
1 & \frac{1}{\sqrt{{\hat c}+|{\hat c}|}}
  & \frac{1}{\sqrt{\hat c}} & \frac{1}{\sqrt{\hat c}} & 1 & 1 
  & \frac{1}{{\hat c}+1} & \frac{1}{{\hat c}+ 1} 
\end{bmatrix},\\
N &=\diag\begin{bmatrix} 
    \frac{{\hat c}}{{\hat c}+1-{\rm i}\delta\arg(\hat c)} 
  & \frac{\hat c}{\sqrt{{\hat c}+|{\hat c}|}}
  & \frac{\sqrt{\hat c}}{2} & \frac{\sqrt{\hat c}}{2} 
  & \frac{{\hat c}}{{\hat c}+|{\hat c}|} & \frac {1}{{\hat\epsilon}+1}
  & 1 & 1
\end{bmatrix},\\
N'&=\diag\begin{bmatrix} 1 & \frac{|{\hat c}|}{\sqrt{{\hat c}+|{\hat c}|}}
  & \sqrt{\hat c} & \sqrt{\hat c} & 1  & 1 & \frac{\hat c}{{\hat c}+1}
  & \frac{\hat c}{{\hat c}+1} 
\end{bmatrix},
\end{split}
\label{eq:3DPN} 
\end{equation} 
where $\hat c=1/\hat k$ and $\hat\epsilon=\hat{k}^2$ as
in~\eqref{eq:kappa}. The matrices $P$ and $N$ differ in their first
elements from those used in~\cite[Theorem 2.3]{HelsRose20}, and we
have set $\hat \mu=1$ as we consider non-magnetic materials. The
modified first elements correspond to another choice of the parameter
$\beta$ in \cite[Section~8]{HelsRose20}, using here $\beta=1+{\rm
  i}\delta\arg(\hat k)$ instead of $\beta=1$, to avoid false
eigenwavenumbers when $(\arg(k_-),\arg(k_+))=(\pi/2,0)$. We use
\begin{align}  
  \delta= 0.2/\pi\,.
\end{align} 
Computations show that this is close enough to $\delta=0$ not to
affect speed and accuracy, but large enough to eliminate false
eigenwavenumbers and near-eigenwavenumbers. In the original
formulation of ``Dirac'' in $\R^3$ from \cite{HelsRose20}, false
eigenwavenumbers appear when $(\arg(k_-),\arg(k_+))=(\pi/2,0)$, that
is, when the wavenumber in the exterior region is imaginary. This
corresponds to the circled lower corner point in
Figure~\ref{fig:amoeba0}(b) and is confirmed by numerical experiments
on the unit sphere and comparison with semi-analytic results given by
Mie theory. For the parameters \eqref{eq:3DPN}, with $\delta=
0.2/\pi$, it follows from \cite[Proposition~8.5]{HelsRose20} that
there are no false eigenwavenumbers for ``Dirac'' in the shaded region
in Figure~\ref{fig:amoeba0}(b). Not even at
$(\arg(k_-),\arg(k_+))=(\pi/2,0)$, where there are true
eigenwavenumbers.

Note that ``Dirac'' is not a Fredholm second kind integral equation
with compact operators on smooth $\Gamma$. For one thing, the block
operator $G$ in~(\ref{eq:G}) contains Cauchy singular
differences of operators. However, the particular choice of $P$, $P'$,
$N$, $N'$ in the original ``Dirac'', that is for $\delta=0$, makes
$G^4$ a compact operator on smooth $\Gamma$ and as a consequence the
spectrum of $G$ has zero as its only accumulation point. This should
be an advantage when using iterative solvers for~(\ref{eq:3Dgen}).

When evaluating $\myvec E$ and $\myvec H$ with ``Dirac'', one can use
either~\eqref{eq:regdens} or~\eqref{eq:tweakeddens} for $h^\pm$. The
reason for preferring \eqref{eq:tweakeddens}, which we use for smooth
$\Gamma$ in the numerical examples of Section~\ref{sec:numex}, is
that, like in~\eqref{eq:maxwrhs}, components $1$ and $5$ of $h^\pm$
from~\eqref{eq:tweakeddens} are zero. This leads to at most five
non-zero densities in the evaluation of each field. However, for
non-smooth $\Gamma$ the numerical method used in
Section~\ref{sec:numex} is less compatible with~\eqref{eq:tweakeddens}
and we use the simpler~\eqref{eq:regdens}. For ``Dirac'', the
densities computed with~\eqref{eq:tweakeddens} satisfy
\begin{equation}
  P h^+=N(f^{\rm in}-h^-),
\end{equation}
so only one of the Cauchy integrals in \eqref{eq:tweakeddens} needs
to be computed.

\subsection{Choice of $P$, $P'$, $N$, $N'$ for ``HK 8-dens''}
\label{sec:HK8dens}

``HK 8-dens'' specifies the diagonal matrices
\begin{equation}
\begin{split}
P &=
{\rm diag}
\begin{bmatrix}
\frac{ \hat k^3\gamma_1\eta}{1+\gamma_1\eta} & 
\frac{-\hat k\lambda\hat\epsilon\eta}{1+\lambda\hat\epsilon\eta} & 
\frac{-\hat k\eta}{1+\eta} & 
\frac{ \hat k\eta}{1+\eta} & 
\frac{ \hat\epsilon\gamma_2\eta}{1+\gamma_2\eta} & 
\frac{-\hat\epsilon\eta}{1+\eta} & 
\frac{ \lambda\hat\epsilon\eta}{1+\lambda\hat\epsilon\eta} & 
\frac{-\lambda\hat\epsilon\eta}{1+\lambda\hat\epsilon\eta}
\end{bmatrix},\\
P'&={\rm diag}
\begin{bmatrix}1&-1&-1&1&1&-1&1&-1
\end{bmatrix},\\
N &={\rm diag}
\begin{bmatrix}
\frac{-1}{1+\gamma_1\eta} & \frac{1}{1+\lambda\hat\epsilon\eta} & 
\frac{1}{1+\eta} & \frac{-1}{1+\eta} & \frac{-1}{1+\gamma_2\eta} & 
\frac{1}{1+\eta} & \frac{-1}{1+\lambda\hat\epsilon\eta} & 
\frac{1}{1+\lambda\hat\epsilon\eta}
\end{bmatrix},\\
N'&={\rm diag}
\begin{bmatrix}
-\hat c^3 & \hat c & \hat c & -\hat c & 
-\hat\epsilon^{-1} & \hat\epsilon^{-1} & -1 & 1
\end{bmatrix},
\end{split}
\label{eq:PNHK}
\end{equation}
where $\gamma_1$, $\gamma_2$, $\eta$, and $\lambda$ are uniqueness
parameters whose determination is discussed in~\cite[Section~11.1 and
Appendix D]{HelsKarl20}. Note that in~\cite{HelsKarl20}, these
parameters are called $\gamma_{\rm E}$, $\gamma_{\rm M}$, $c$, and
$\lambda$. A valid choice when $(\arg(k_-),\arg(k_+))=(0,0)$ is
\begin{equation}
\begin{bmatrix}
\gamma_1 & \gamma_2 & \eta & \lambda
\end{bmatrix}=
\begin{bmatrix}
\hat\epsilon^{-1} & 1 & 1 & 1
\end{bmatrix}.
\label{eq:posreal}
\end{equation}
This is also a valid choice in the part of the uniqueness region of
Figure~\ref{fig:amoeba0}(b) that is in the vicinity of
$(\arg(k_-),\arg(k_+))=(\pi/2,0)$. For this reason \eqref{eq:posreal}
is used also at $(\arg(k_-),\arg(k_+))=(\pi/2,0)$. A valid choice when
$(\arg(k_-),\arg(k_+))=(0,\pi/2)$ is
\begin{equation}
\begin{bmatrix}
\gamma_1 & \gamma_2 & \eta & \lambda
\end{bmatrix}=
\begin{bmatrix}
{\rm i}\hat\epsilon^{-1} & 1 & -{\rm i} & {\rm i}
\end{bmatrix}.
\label{eq:plasmonic}
\end{equation} 

The surface densities $h$ of~(\ref{eq:h}) have, with ``HK 8-dens'', the
physical interpretations
\begin{multline}
\begin{bmatrix}
h_1 & h_2 & h_3 & h_4 & h_5 & h_6 & h_7 & h_8
\end{bmatrix}
=\\
\begin{bmatrix}
\sigma_{\rm E} & \varrho_{\rm M} & J_\theta & J_\tau & 
\sigma_{\rm M} & \varrho_{\rm E} & M_\theta & M_\tau
\end{bmatrix},
\label{eq:physint}
\end{multline}
where $-{\rm i}k_-\sigma_{\rm E}$ and $-{\rm i}k_-\sigma_{\rm M}$ are
exterior limits of the electric and magnetic volume charge densities
on $\Gamma$, $\varrho_{\rm E}$ and $\varrho_{\rm M}$ are the
equivalent electric and magnetic surface charge densities on the
exterior side of $\Gamma$, and $J_\theta$, $J_\tau$, $M_\theta$,
$M_\tau$ are components of the equivalent electric and magnetic
surface current densities on the exterior side of $\Gamma$.
See~\cite[Remark~10.2]{HelsKarl20}, where it also is shown that
$h_1=\sigma_{\rm E}$ and $h_5=\sigma_{\rm M}$ must be zero on
theoretical grounds. Therefore, the preferred choice for field
evaluations with ``HK 8-dens'' is always via \eqref{eq:regdens}, which
is the Stratton--Chu representation, and \eqref{eq:tweakeddens} is
never needed since it does not lead to any reduced numerical costs.
   
We remark that ``HK 8-dens'' was discovered independently and prior to
``Dirac'', and that it was only later realized that it can be written
in the form \eqref{eq:3Dgen}--\eqref{eq:G}, just like ``Dirac''.
However, it is not the case that ``HK 8-dens'' is a special case of
``Dirac'', corresponding to a certain choice of Dirac parameters as in
\cite[Section 8]{HelsRose20}. Indeed, ``HK 8-dens'' is not derivable
from jump matrices $M$ and $M'$ as in \cite[Theorem 2.3]{HelsRose20}.

\section{False essential spectra}
\label{sec:ess}

A key operator for the MTP($k_-,k_+$) is the static, $k\to 0$, limit
\begin{equation}
     K_{\rm d} g(\myvec r)=
\text{\rm p.v.\!}
\int_\Gamma \myvec\nu(\myvec r')\cdot\nabla'\Phi_0(\myvec r,\myvec r') 
g(\myvec r')\,{\rm d}\Gamma'\,,
\quad\myvec r\in\Gamma,
\end{equation}
of the acoustic double layer operator $K^{\myvec\nu'}_k$, appearing in
the $(1,1)$ and $(5,5)$ diagonal blocks of $E_k$. That is, $K_{\rm d}$
equals the Neumann--Poincar\'e operator $K_{\rm NP}$, possibly modulo
a sign depending on convention~\cite{Ando20}. Its essential spectrum
$\sigma_{\text{ess}}(K_{\rm d})$ in the fractional Sobolev space
$H^{1/2}(\Gamma)$, that is the set of $\lambda$ for which $\lambda
I-K_{\rm d}$ fails to be a Fredholm operator, is a compact subset of
the interval $(-1,1)$, for any Lipschitz surface $\Gamma$. Unlike in
$\R^2$, $\sigma_{\text{ess}}(K_{\rm d})$ is not necessarily symmetric
with respect to $0$ for $\Gamma\subset\R^3$: see examples for the
spectrum of the adjoint of $K_{\rm NP}$, $\sigma_{\text{ess}}(K_{\rm
  NP}^*)$ in $H^{-1/2}(\Gamma)$, and $\Gamma$ with axially symmetric
conical points in~\cite[Section 7.3]{HelsPerf18}.

We map $\sigma_{\text{ess}}(K_{\rm d})$ onto
the negative real axis and define
\begin{align}
  \Sigma(\Gamma)=\sett{\hat\epsilon\in\C}
  {(1+\hat \epsilon)/(1-\hat\epsilon)\in \sigma_{\text{ess}}(K_{\rm d})}. 
\end{align}
For $E_k$ and our IERs, the relevant function space is
\begin{multline}   
  \label{eq:H3space}
  \mH_3 = 
  H^{1/2}(\Gamma)\oplus
  H^{-1/2}(\Gamma)\oplus
  H^{-1/2}(\curl,\Gamma) \\
  \qquad\oplus 
  H^{1/2}(\Gamma)\oplus
  H^{-1/2}(\Gamma)\oplus
  H^{-1/2}(\curl,\Gamma),
\end{multline}
as discussed in \cite[Section~5]{HelsRose20}. Here
$H^{-1/2}(\curl,\Gamma)$ denotes the tangential vector fields in
$H^{-1/2}$ with tangential $\curl$ in $H^{-1/2}$, with suitable
modification for non-smooth $\Gamma$. By inspection of the proof of
\cite[Proposition 8.4]{HelsRose20} it is immediate that the ``Dirac''
IER is a Fredholm operator in $\mH_3$ if and only if the
MTP($k_-,k_+$) defines a Fredholm map. By the latter we mean that
$(\myvec E^{\rm tr},\myvec H^{\rm tr},\myvec E^{\rm sc},\myvec H^{\rm
  sc}) \mapsto f^{\rm in}$ is a Fredholm map, in $L^2$-norms of the
fields in a bounded neighborhood of $\Gamma$ in $\R^3$ and $f^{\rm
  in}\in\mH_3$ being the trace of a Maxwell field as in
\eqref{eq:maxwrhs}. Furthermore, as we shall prove in a forthcoming
publication, the MTP($k_-,k_+$) defines a Fredholm map if and only if
$\hat\epsilon\notin\Sigma(\Gamma)$, for any wavenumbers and not only
in the quasistatic limit.

Another key operator for the MTP($k_-,k_+$) is the magnetic dipole
operator,
\begin{equation}
     K_{\rm m} g(\myvec r)=\myvec\nu(\myvec r)\times
\text{\rm p.v.\!}
\int_\Gamma \nabla'\Phi_0(\myvec r,\myvec r') \times
g(\myvec r')\,{\rm d}\Gamma'\,,
\quad\myvec r\in\Gamma,
\end{equation}
acting on tangential vector fields $g$. We note that the static limit
of the operator appearing in the (3:4,3:4) and (7:8,7:8) size $2\times
2$ diagonal blocks in $E_k$, is $-K_{\rm m}^*$. Moreover, the static
limit of the normal derivative of the acoustic single layer potential,
$K^{\myvec\nu}_k$, appearing in the $(2,2)$ and $(6,6)$ diagonal
elements, equals $-K_{\rm d}^*$.

By Hodge decomposition of $H^{-1/2}(\curl,\Gamma)$, it can be shown
that the essential spectrum of $K_{\rm m}$ is
$\sigma_{\text{ess}}(K_{\rm d})\cup(-\sigma_{\text{ess}}(K_{\rm d}))$.
The corresponding result for eigenvalues is
in~\cite[Proposition~4.7]{Ammari16} and in~\cite{Kress95}. However,
these results are proved for smooth $\Gamma$, on which $K_{\rm m}$ is
compact and $\sigma_{\text{ess}}(K_{\rm d})=\{0\}$.

In the diagonal blocks of the matrix $E_k$, we therefore find
operators with essential spectra $\sigma_{\text{ess}}(K_{\rm d})$ as
well as $-\sigma_{\text{ess}}(K_{\rm d})$. To avoid the {\em false
  essential spectrum} $(-\sigma_{\text{ess}}(K_{\rm d}))\setminus
\sigma_{\text{ess}}(K_{\rm d})$, the matrices $P,P',N,N'$ need to be
chosen carefully. Consider the diagonal operator blocks
of~(\ref{eq:Ek3D}) for ``Dirac'' and ``HK 8-dens'' when $\hat\epsilon$
is negative real and $\hat k/{\rm i}$ is positive, referred to as the
plasmonic case in Section~\ref{sec:numex}. The only operator block
which can fail to be a Fredholm operator for ``Dirac'' is (6,6). This
happens when $\hat\epsilon\in\Sigma(\Gamma)$, that is, if and only if
the MTP($k_-,k_+$) itself fails to define a Fredholm map.

The diagonal operator blocks of ``HK 8-dens'' that can fail to be
Fredholm operators are the (1,1), (2,2) and (7:8,7:8) blocks. The
(1,1) and (2,2) blocks each fails to be a Fredholm operator when
$\hat\epsilon^{-1}\in\Sigma(\Gamma)$, whereas the (7:8,7:8) size
$2\times 2$ diagonal block fails to be a Fredholm operator when
$\hat\epsilon\in\Sigma(\Gamma)$ or
$\hat\epsilon^{-1}\in\Sigma(\Gamma)$. However, in the analysis of the
full IERs also the non-diagonal blocks need to be taken into account,
as the plots of densities in Section~\ref{sec:numex} clearly show.

The spectral properties of the diagonal blocks show that ``Dirac'' has
no false essential spectrum, but also indicate that ``HK 8-dens'' may
have false essential spectrum. Section~\ref{sec:tomato} contains
numerical results that support this. However, since the space $\mH_3$
is a space of mixed $\pm 1/2$ regularity, a full proof must include an
analysis of the off-diagonal blocks. We plan a forthcoming publication
devoted to a more careful theoretical and numerical study of the
essential spectrum and the issues discussed in this section.

\section{The quasi-static limit} 
\label{sec:static}

In the quasi-static limit $k_\pm\to 0$, with $\hat k= k_+/k_-$
fixed, the operator $G$ from $\eqref{eq:G}$ simplifies
considerably. The diagonal matrices $P,P',N, N'$ are all fixed
whereas our basic Cauchy integral operator becomes a $4/4$ block
diagonal operator
\begin{equation}
E_0=
\begin{bmatrix}
-K_{\rm d}  & 0 & {\myvec K_{1,3:4}} &
 0 & 0 & {\myvec 0}  \\
 K^{\myvec\nu\times\myvec\nu'}_0 & K_{\rm d}^* & 
  {\myvec K_{2,3:4}} & 0 & 0 & {\myvec 0} \\
 {\myvec K_{3:4,1}} & {\myvec K_{3:4,2}} & 
-{\myvec K_{\rm m}^*} & {\myvec 0} & {\myvec 0} & {\myvec 0}\\
  0 & 0 & {\myvec 0} & -K_{\rm d} & 0 & {\myvec K_{5,7:8}} \\
 0 & 0 & {\myvec 0} & -K^{\myvec\nu\times\myvec\nu'}_0 &
 K_{\rm d}^* & {\myvec K_{6,7:8}} \\
 {\myvec 0} & {\myvec 0} & {\myvec 0} & {\myvec K_{7:8,5}} & 
{\myvec K_{7:8,6}} & -{\myvec K_{\rm m}^*}
\end{bmatrix},
\label{eq:Ek3Dstat}
\end{equation}
where we have written rows/columns 3:4 and 7:8 in vector/block
notation. We keep indexing 1:8, and all operators are as in
\eqref{eq:Ek3D} but with $k=0$. Now all single layer operator entries
vanish, as they contain a factor $k_\pm$, and the equations for the
magnetic and electric fields, which correspond to the two diagonal
blocks, decouple. For spectral properties of the operators $K_{\rm d}$
and $K_{\rm m}$ defined in Section~\ref{sec:ess}, used in this
section, we refer to~\cite[Sections 5.1--5.2]{ColtonKress92}. These
results generalize to Lipschitz surfaces, if we use the spaces
$H^{1/2}(\Gamma)$ and $H^{-1/2}({\rm curl},\Gamma)$ for $K_{\rm d}$
and $K_{\rm m}$ respectively.

Let MTP($0,0,\hat k$) denote the Maxwell transmission problem in the
quasi-static limit with $\hat k$ fixed. The MTP($0,0,\hat k$) amounts
to two decoupled divergence and curl free vector fields $\myvec E$ and
$\myvec H$ in $\Omega_\pm$, having continuous tangential parts
\eqref{eq:rv2C}--\eqref{eq:rv1C} and with $\myvec E^{\rm sc}$ and
$\myvec H^{\rm sc}$ decaying at infinity. We now also explicitly need
to require the Gauss jump conditions
\begin{align}
\lim_{\Omega_-\ni\myvec r'\to\myvec r}
\myvec\nu\cdot\myvec E(\myvec r')&=
\hat\epsilon \lim_{\Omega_+\ni\myvec r'\to\myvec r}
\myvec\nu\cdot\myvec E(\myvec r'),
\quad\myvec r\in\Gamma,\\
\lim_{\Omega_-\ni\myvec r'\to\myvec r}
\myvec\nu\cdot\myvec H(\myvec r')&=
\lim_{\Omega_+\ni\myvec r'\to\myvec r}
\myvec\nu\cdot\myvec H(\myvec r'),
\quad\myvec r\in\Gamma.
\end{align} 
We note that since we consider non-magnetic materials, the jump
condition for all components of $\myvec H$ are the same. It is only
the (5:8,5:8) diagonal block in \eqref{eq:Ek3Dstat} which needs to be
inverted in the quasi-static limit. Moreover, one can show that the
MTP($0,0,\hat k$) defines an invertible map if and only if
$(1+\hat\epsilon)/(1-\hat\epsilon)\notin\sigma(K_{\rm d})$.

In Section~\ref{sec:staticplots} we show numerical results for
condition numbers of ``Dirac'' as well as of ``HK 8-dens'' in this
quasi-static limit, but restrict our analysis here to ``Dirac''. For
simplicity, assume $\delta=0$ in the definition of $P,P',N,N'$
in~(\ref{eq:3DPN}). This makes the blocks in $G$ corresponding to
$K_{1,3:4}$, $K_{2,3:4}$, $K_{7:8,5}$ and $K_{7:8,6}$ vanish, and
invertibility of $I+G$ is determined by the diagonal blocks.
($\delta\approx 0$ makes $K_{1,3:4}\approx 0$.) The main operator is
the (6,6) diagonal block, which fails to be invertible for ``Dirac''
precisely when $(1+\hat\epsilon)/(1-\hat\epsilon)\in\sigma(K_{\rm
  d})$. Since the spectra of $K_{\rm d}$, $K_{\rm d}^*$ and $K_{\rm
  m}^*$ are subsets of $[-1,1]$, for any topology of $\Gamma$, the
remaining diagonal operators are all seen to be invertible for any
$\hat k\ne 0$ and $\hat k\ne\infty$.

Let MTP($0,0,0$) and MTP($0,0,\infty$) denote the MTP($0,0,\hat k$) in
the limits $\hat k\to 0,\infty$. The MTP($0,0,0$) can be viewed as an
exterior homogeneous Neumann problem for the scalar electric potential
and the MTP($0,0,\infty$) as an exterior Dirichlet problem. Both
problems can be shown to have unique solutions. To analyze the
behavior of ``Dirac'', with $\delta=0$ as above, for these problems,
we examine which spectral points are used for the diagonal blocks in
$I+G$. From \cite[Equation (132)]{HelsRose20}, we see that for the
diagonal (2,2), (3:4,3:4) and (5:5) blocks, the spectral points are
uniformly bounded away from $[-1,1]$ as $\hat k\to 0,\infty$. The
(6,6) block is $I-\tfrac{1-\hat\epsilon}{1+\hat\epsilon}K_{\rm d}^*$
as discussed above. The (1,1) block is
\begin{equation}
  I-\tfrac{\hat k-1}{\hat k+1}K_{\rm d}
\end{equation}
and the (7:8,7:8) block is 
\begin{equation}
  I+\tfrac{\hat k-1}{\hat k+1}K_{\rm m}^*.
\end{equation}
Here $\sigma(K_{\rm d})\cap\{-1,1\}=\{-1\}$. Furthermore
$\sigma(K_{\rm m}^*)\cap\{-1,1\}=\emptyset$, assuming that $\Omega_+$
is simply connected. This shows that ``Dirac'' exhibits a false
eigenwavenumber for the MTP($0,0,0$), due to the (1,1) block. If one
is only interested in the quasi-static limit, this deficiency can be
avoided by only using the (5:8,5:8) block of $I+G$ for solving for the
electric fields as discussed above. (Or even simpler IERs available in
this case.) If one is also interested in near quasi-statics and $\hat
k\approx 0$, assuming that $\Omega_+$ is simply connected, we can tune
the free Dirac parameters $r,\beta,\gamma,\alpha',\beta',\gamma'$ from
\cite[Section~8]{HelsRose20} as follows. Rather than choosing
$\alpha'=\beta'=1/\hat k$ as done for ``Dirac'', we choose
$\alpha'=|\hat k|/\hat k$ and $\beta'= 1/(|\hat k|\hat k)$. With these
choices, the spectral point for the (1,1) block stays uniformly
bounded away from $[-1,1]$. We approach the spectral point $+1$ for
the (6,6) and (7:8,7:8) blocks as $\hat k\to 0$, but this is not a
problem since $+1\notin\sigma(K^*_{\rm d})$ and, for simply connected
$\Omega_+$, $+1\notin \sigma(K_{\rm m}^*)$. Preconditioning similarly
to \cite[Theorem~2.3]{HelsRose20}, we arrive at
\begin{equation}
\begin{split}
  P &= \begin{bmatrix} \frac{\beta\sqrt{|\hat c|/\hat c}}{{\hat c}+|\hat c|\beta} & \frac {\hat c^{-1}}{\sqrt{1+|{\hat c}|/\hat c}}
  & \frac 1{2\hat c} & \frac 1{2\hat c} & \frac {|{\hat c}|}{{\hat c}+ |{\hat c}|} & \hat \epsilon & \frac {|\hat c|\hat c}{1+|\hat c|\hat c} & 
  \frac {|\hat c|\hat c}{1+|\hat c|\hat c} \end{bmatrix},\\
  P' &= \begin{bmatrix} \sqrt {\frac {\hat c}{|\hat c|}} & 
 \frac 1{\sqrt{1+|{\hat c}|/\hat c}}
  & 1 & 1 & 1 & \frac 1{\hat\epsilon+1} & \frac 1{|\hat c|\hat c} & \frac 1{|\hat c|\hat c}
     \end{bmatrix},  \\
  N &= \begin{bmatrix} \frac{\sqrt{|\hat c|/\hat c}}{1+\beta|\hat c|/\hat c} & \frac 1{\sqrt{1+|{\hat c}|/\hat c}}
  & \frac 12 & \frac 12 & \frac {{\hat c}}{{\hat c}+ |{\hat c}|}  & 1 &  \frac {|\hat c|\hat c}{1+|\hat c|\hat c} & \frac {|\hat c|\hat c}{1+|\hat c|\hat c}   \end{bmatrix},  \\
  N' &= \begin{bmatrix} \sqrt {|\hat c|\hat c} & \frac {|{\hat c}|}{\sqrt{1+|{\hat c}|/\hat c}}
  & \hat c & \hat c & 1  & \frac 1{\hat\epsilon+1} & 1 & 1 
    \end{bmatrix}, 
\end{split}
\label{eq:pecdir} 
\end{equation}
where $\beta=1-{\rm i}\delta\arg(\hat c)$ is chosen as in
Section~\ref{sec:Dirac}. When $k_\pm\approx 0$ and $\hat k\approx 0$,
the version~(\ref{eq:pecdir}) of ``Dirac'' is better conditioned than
that of~(\ref{eq:3DPN}). Indeed, one verifies for these choices of
parameters and $P,P'$ that $I+G$ converges in operator norm as
$k_\pm\to 0$ and $\hat k\to 0$, to a limit operator which is
invertible due to the block diagonal and triangular structures and the
fact that the (1,1) block is now invertible as above.

For the MTP($0,0,\infty$), ``Dirac'' again has a false
eigenwavenumber, this time due the (6,6) block. It is not clear how to
avoid this by tuning the free Dirac parameters. We plan to address
this issue, as well as $\Omega_+$ that are not simply connected, in a
forthcoming paper.

\section{Surface plasmon standing waves and plasmons}
\label{sec:plasmons}

Let, for the moment, $\R^3$ be divided into two, not necessarily
bounded, domains $\Omega_1$ and $\Omega_2$ separated by a surface
$\Gamma$ and with real-valued permittivities $\epsilon_1$ and
$\epsilon_2$ such that $\epsilon_1>0$ and $\epsilon_2<0$. {\em Surface
  plasmon waves}, {\em surface plasmon standing waves}, and {\em
  quasi-static plasmons} are then particular types of electromagnetic
fields that may appear as solutions to the MTP($k_-,k_+$). These
fields are briefly described here from a classical electrodynamics
point of view and they are referenced in the discussion of the
numerical examples in Section~\ref{sec:numex}.

Surface plasmon waves (SPWs) are surface waves that can propagate
along $\Gamma$. A necessary extra condition for their existence on
planar $\Gamma$ is $\hat\epsilon\equiv\epsilon_2/\epsilon_1<-1$, see
\cite[Section~5.1]{SihvQiLind10} and \cite[Appendix~I]{Raet88}.
Numerical experiments done in the preparation of \cite{HelsKarl20}
indicate that the same condition holds on non-planar $\Gamma$. On
planar $\Gamma$, the SPWs propagate without attenuation with
wavelength
\begin{equation}
\lambda_{\rm spw}=
\lambda_1\sqrt{1-1/\vert \hat\epsilon\vert}
\label{eq:SPW}
\end{equation}
and decay exponentially in the directions normal to $\Gamma$, see
\cite[Section~2.1]{SihvQiLind10} and~\cite[Appendix I]{Raet88}. Here
$\lambda_1=2\pi/k_1$ is the free space wavelength in $\Omega_1$.
Furthermore, the following has been verified by us for a circular
cylinder, using semi-analytic methods, and by numerical examples in
\cite{HelsKarl18,HelsKarl20}: The SPWs can propagate without
attenuation along surfaces that are invariant only in the propagation
direction. They can also propagate along surfaces which are curved in
the direction of propagation, but are then attenuated due to
radiation. The radiation increases with the ratio of $\lambda_{\rm
  spw}$ to the radius of curvature.

A natural conjecture is that SPWs appear, excited at corners or edges
on $\Gamma$, when $(1+\hat\epsilon)/(1-\hat\epsilon)$ hits the
essential spectrum of $K_{\rm d}$. This was discussed, for edges,
in~\cite[Section 7.3]{HelsKarl18}, but that discussion is not
exhaustive. Indeed, as we shall see in Section~\ref{sec:numex}, an SPW
appears in Figure~\ref{fig:tomato} even though we are outside the
essential spectrum, and in Figure~\ref{fig:drop} we are in the
essential spectrum, but no SPW appears. The description of the precise
mechanisms for the excitation of SPWs on non-smooth $\Gamma$ remains
an open problem.
   
Consider now, with notation as in Section~\ref{sec:probform}, a
bounded object $\Omega_+$ and assume that the wavenumber $k_-$ is
positive real and that $\hat{\epsilon}<-1$, to enable SPWs along
$\Gamma$. At certain $k_-$ the SPWs form standing waves. Such SPWs are
here called surface plasmon standing waves (SPSWs). In some literature
the name surface plasmon resonances (SPRs) is used to emphasize their
resonant nature. However, SPR is not a direct synonym for SPSW since
it is also used for other plasmonic phenomena. Due to radiation, the
SPSW is a damped resonant field and the $(k_-,k_+)$, for which it
appears, is close to a true eigenwavenumber outside the set of points
shown in Figure~\ref{fig:amoeba0}(b). That is, SPSWs appear close to
true near-eigenwavenumbers, in the terminology from
Section~\ref{sec:prop}. When an SPSW is excited by an incident field,
its amplitude becomes large, which results in a large scattering cross
section. This is one reason why SPSWs are of interest in optics.

A quasi-static plasmon is an electromagnetic field that approaches an
eigenfield, referred to as a {\em static plasmon}, as $k_\pm\to 0$. It
appears around objects that are much smaller than the wavelength
$\lambda_-$. The quasi-static plasmon is also denoted surface plasmon
(SP) in optics, a name that may lead to confusion since it is
sometimes used as a synonym for SPSW. A quasi-static plasmon is a
resonant field, but in contrast to SPSWs it is not a standing wave
field. For smooth $\Gamma$ there is an infinite discrete set of
$\hat\epsilon$ corresponding to static plasmons. The electric field of
a static plasmon is distributed so that the electric energies in
$\Omega_-$ and $\Omega_+$ exactly cancel each other, while the
magnetic field is zero. Indeed, Green's identity shows that
$\hat\epsilon\int_{\Omega_+}|\myvec E|^2 dx= -\int_{\Omega_-}|\myvec
E|^2 dx$ so that, in accordance with Poincar\'e's variational
principle \cite[Theorem~2.4]{Ando20}, the static plasmon corresponds
to the eigenvalue $(1+\hat\epsilon)/(1-\hat\epsilon)$ for $K_{\rm d}$.

Quasi-static and static plasmons can be classified as being {\em
  bright} or {\em dark} depending on whether they can be excited by a
uniform incident field or not. A bright plasmon radiates as an
electric dipole and its far-field is very large considering the object
is small compared to $\lambda_-$. A dark plasmon radiates as a
higher-order multipole and its far-field is weak. See \cite[Lemma
5.3]{Ammari16}, where the full multipole expansion of far-fields is
given. Furthermore, there is a close connection between bright static
plasmons and the imaginary part of an object's limit
polarizability~\cite{HelsPerf13,Perf20}. For an object with sharp
corners, edges, or points there is a continuous, possibly punctured,
interval of $\hat\epsilon$ where a special type of static plasmons,
with infinite (but cancelling) electric energies in $\Omega_-$ and
$\Omega_+$, can occur together with a partially embedded discrete set
of regular static plasmons~\cite{HelsKangLim17,HelsPerf18,Perf20}.

\section{Scattering objects and discretization}

This section reviews shapes of scattering objects and discretization
schemes that are used for the numerical tests in
Section~\ref{sec:numex}.

\subsection{Two surface families}

Examples with smooth $\Gamma$ come from a generating curve $\gamma$
parameterized as
\begin{equation}
r(s)=(1+\alpha\sin(5s))(\cos(s),\sin(s))\,,\quad s\in[-\pi/2,\pi/2]\,,
\label{eq:starfish}
\end{equation}
where $\alpha$ is a shape parameter. The choice $\alpha=0$ corresponds
to $\Gamma$ being the unit sphere. With $\alpha=0.25$ the shape of
$\Gamma$ resembles a ``rotated starfish''. See
Figure~\ref{fig:notation}(a).

Examples with non-smooth $\Gamma$ come from a $\gamma$ parameterized
as
\begin{equation}
r(s)=\sin(\pi s)\left(\sin((0.5-s)\alpha),\cos((0.5-s)\alpha)\right)\,,
\quad s\in[0,0.5]\,,
\label{eq:tomato}
\end{equation}
where $\alpha$ is the opening angle of the conical point at the
origin. With $\alpha>\pi$, the shape of $\Gamma$ resembles a
``tomato''. See Figure~\ref{fig:amoeba0}(a) for an illustration with
$\alpha=31\pi/18$. With $\alpha<\pi$ the shape of $\Gamma$ resembles a
``drop with a sharp tip''. The generating curve $\gamma$
of~(\ref{eq:tomato}) has previously been used for numerical examples
in~\cite{HelsKarl20,HelsPerf18}.

\subsection{Fourier--Nyström discretization}  
\label{sec:FouNysd} 

The integral equation~(\ref{eq:3Dgen}) is solved on axially symmetric
$\Gamma$ using high-order Fourier--Nyström discretization. An
azimuthal Fourier transform is first applied to~(\ref{eq:3Dgen}),
yielding a sequence of modal problems for the Fourier coefficients
$h_{(n)}$ of $h$
\begin{equation}
  (I+G_{(n)})h_{(n)}(r)= 2Nf^{\rm in}_{(n)}(r)\,,\quad r\in\gamma\,,
\quad n=0,\pm 1,\pm 2,\ldots\,,
\label{eq:3DgenF}
\end{equation}
which are solved independently using high-order panel-based Nyström
discretization. The modal solutions $h_{(n)}$ are then synthesized to
give the full solution $h$.

High-order panel-based Fourier--Nyström discretization for solving
integral equations modeling scattering problems on axisymmetric
surfaces was made popular by Young, Hao, and Martinsson in
2012~\cite{YouHaoMar12}. Since then, several authors have worked on
extensions of such schemes and related issues. The aim has been to
include a broader range of integral operators, to reach higher
achievable accuracy, to evaluate near fields more efficiently, and to
cope with problems related to wavenumbers with large imaginary
parts~\cite{EpsGreNei19,LaiOneil19,HelsKarl14,KliFryTor19}. Our
version of the Fourier--Nyström scheme is the one used
in~\cite{HelsKarl20}. This version is of $16$th order. It relies on a
combination of 16-point and 32-point underlying Gauss--Legendre
quadrature, a variety of explicit kernel-splits, semi-analytic product
integration computed on the fly, and is compatible with the
recursively compressed inverse preconditioning method
(RCIP)~\cite{Helsing18}. The RCIP accelerates and stabilizes Nyström
discretization in the presense of singular boundary points on
$\gamma$, such as corners.

\section{Numerical examples}
\label{sec:numex}

The numerical efficiency of ``Dirac'' and ``HK 8-dens'' is now
compared. That is, we solve the discretized modal
systems~(\ref{eq:3DgenF}) and compute fields
via~(\ref{eq:Efieldminus})--(\ref{eq:Hfieldplus}) with $P$, $P'$, $N$,
$N'$ as in Sections~\ref{sec:Dirac} and~\ref{sec:HK8dens}. We also
compute condition numbers of the modal systems. Three material
parameter cases are used:
\begin{itemize}
\item {\it the positive dielectric case}, where $\hat k=1.5$ and $k_-$
  is positive real;
\item {\it the plasmonic case}, where $\hat k={\rm i}\sqrt{1.1838}$
  and $k_-$ is positive real;
\item {\it the reverse plasmonic case}, where $\hat k=\left({\rm
      i}\sqrt{1.1838}\right)^{-1}$ and $k_+$ is positive real.
\end{itemize}
These parameter cases are taken from previous work on time-harmonic
transmission problems~\cite{Annsop16,HelsKarl20,HelsRose20}. 

In all examples involving field images, unless otherwise stated, the
incident field is a linearly polarized plane wave $\myvec E^{\rm
  in}(\myvec r)={\myvec x}e^{{\rm i}k_-z}$ with ${\myvec x}=(1,0,0)$.
The fields are plotted in the plane $y=0$, where the field components
$E_\theta$, $H_\rho$, and $H_z$ are zero due to symmetry. To save
space in the examples, we only show $E_\rho$ and $H_\theta$.
Generally, these two seem to exhibit more pronounced field patterns
than the omitted component $E_z$.

When $\Gamma$ is non-smooth, it may happen that $\hat k={\rm
  i}\sqrt{1.1838}$ or $\hat k=\left({\rm i}\sqrt{1.1838}\right)^{-1}$
correspond to that the MTP($k_-,k_+$) does not define a Fredholm map.
We then compute limit solutions as $\hat\epsilon$ approaches $-1.1838$
or $-1/1.1838$ from above in the complex plane. Such limit solutions
have boundary traces lying outside the function space $\mH_3$ from
\eqref{eq:H3space} and are given a downarrow superscript. For example,
the limit of the field $\myvec E$ is denoted $\myvec E^\downarrow$.

Our codes are implemented in {\sc Matlab}, release 2018b, and executed
on a workstation equipped with an Intel Core i7-3930K CPU and 64 GB of
RAM. Large linear systems are solved iteratively using GMRES with
a stopping criterion threshold of machine epsilon ($\epsilon_{\rm
  mach}$) in the estimated relative residual. When assessing the
accuracy of computed field quantities we adopt a procedure where to
each numerical solution we also compute an overresolved reference
solution, using roughly 50\% more points in the discretization of the
system under study. The absolute difference between these two
solutions is denoted the {\it estimated absolute error}.
In examples involving field images, the fields are computed at
$10^6$ target points on a rectangular Cartesian grid in the
computational domains shown.
 
Subsections~\ref{sec:sphere}, \ref{sec:convergence},
and~\ref{sec:staticplots}, on eigenwavenumbers, convergence, and the
quasi-static limit, involve the unit sphere. An advantage with doing
experiments on spheres is that semi-analytic results given by Mie
theory can be used for verification. The remaining subsections concern
less trivial object shapes and there we limit ourselves mainly to the
plasmonic case. The positive dielectric case is not deemed challenging
enough for thorough testing. The performances of ``Dirac'' and ``HK
8-dens'' in the reverse plasmonic case is rather similar to their
respective performances in the plasmonic case -- apart from the new
risk of being close to a true eigenwavenumber and, for ``HK 8-dens'',
also close to a false eigenwavenumber or near-eigenwavenumber.

\begin{figure}[t!]
\centering
\includegraphics[height=44mm]{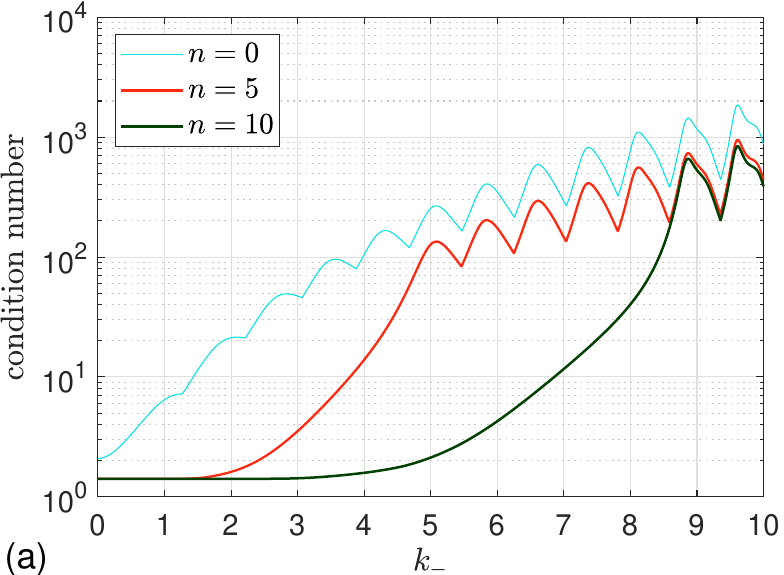}
\hspace*{2mm}
\includegraphics[height=44mm]{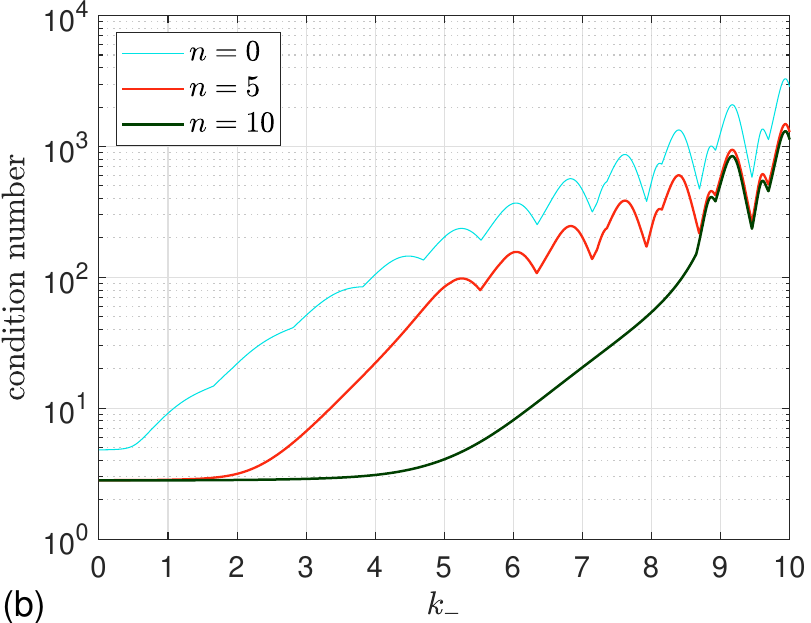}

\vspace{3mm}
\includegraphics[height=44mm]{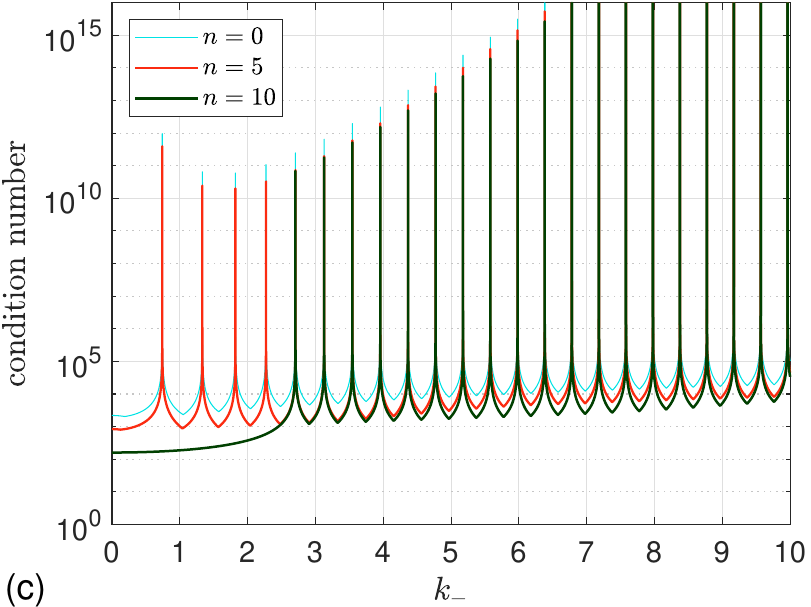}
\hspace*{1mm}
\includegraphics[height=44mm]{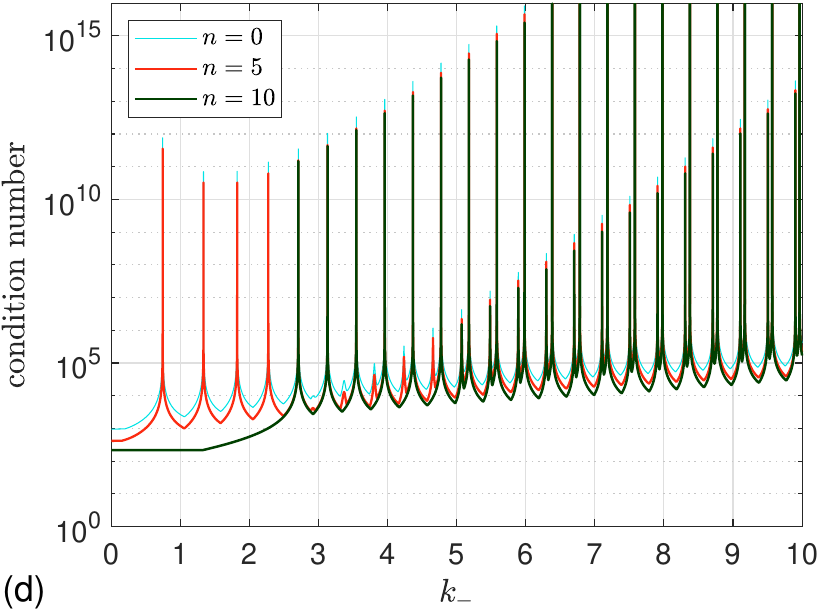}

\vspace{3mm}
\includegraphics[height=44mm]{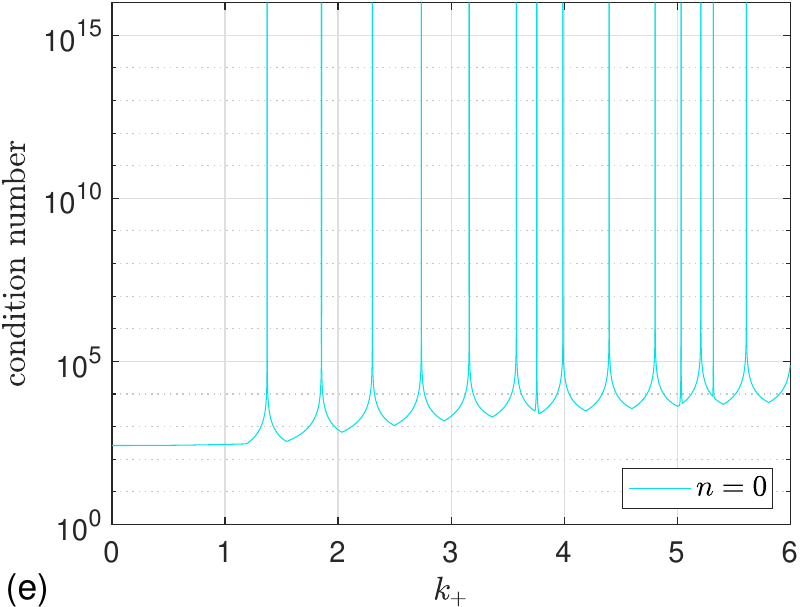}
\hspace*{2mm}
\includegraphics[height=44mm]{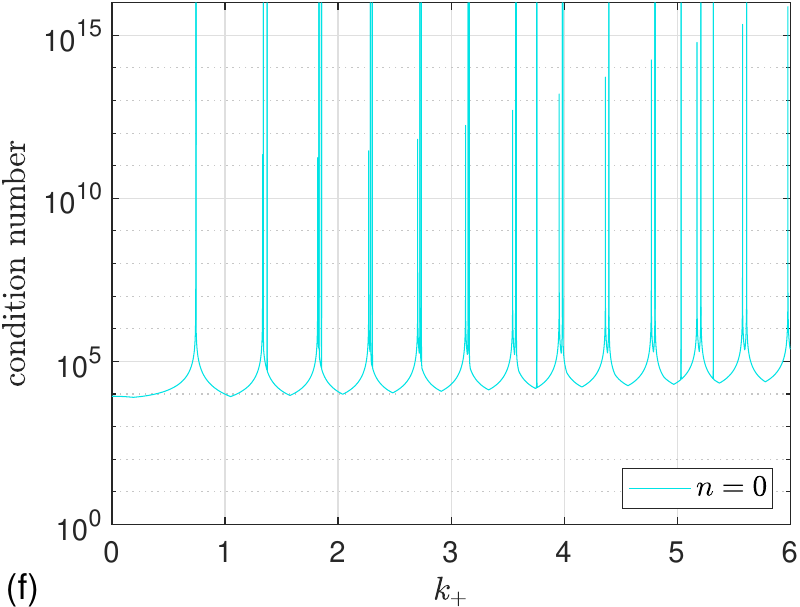}
\caption{\sf The unit sphere. Condition numbers of system matrices of 
  the discretized modal integral equations~(\ref{eq:3DgenF}) with
  azimuthal indices $n=0,5,10$ and wavenumber $k_-\in[0,10]$ or
  $k_+\in[0,6]$: (a,b) the positive dielectric case with $\hat
  k=1.5$; (c,d) the plasmonic case with $\hat k={\rm i}\sqrt{1.1838}$;
  (e,f) the reverse plasmonic case with $\hat k=\left({\rm
      i}\sqrt{1.1838}\right)^{-1}$; (a,c,e) ``Dirac''; (b,d,f) ``HK
  8-dens''.}
\label{fig:sphere}
\end{figure}

\subsection{Condition numbers on the unit sphere}
\label{sec:sphere}

We compute condition numbers of system matrices of the
discretized modal integral equations~(\ref{eq:3DgenF}) and with
$\Gamma$ being the unit sphere. Results for the azimuthal modes
$n=0,5,10$ are shown in Figure~\ref{fig:sphere}. A number of $768$
discretization points are placed on a grid on $\gamma$, making the
total system size $6144\times 6144$, and up to $5,\!300$ data
points are used to capture the rapid variations in the condition
numbers as $k_-$ and $k_+$ are swept through the intervals
$[0,10]$ and $[0,6]$. The purpose of this study is primarily to detect
possible false eigenwavenumbers, but also to get a notion of how
well-conditioned the two IERs under study are.

Figure~\ref{fig:sphere} shows that ``Dirac'' and ``HK 8-dens'' have
similar condition numbers in the positive dielectric case, while
``Dirac'' is better conditioned than ``HK 8-dens'' in the plasmonic-
and reverse plasmonic cases -- particularly at higher wavenumbers
$k_-$ and $k_+$. Note that the regularly recurring high peaks that are
common to Figures~\ref{fig:sphere}(c,d) correspond to true
near-eigenwavenumbers, each with multiplicity one for a given $n$,
while the additional peaks in Figure~\ref{fig:sphere}(d) correspond to
false near-eigenwavenumbers of ``HK 8-dens'', as discussed in the last
paragraph of Section~\ref{sec:prop}. The true near-eigenwavenumbers
are associated with SPSWs. The first peak in
Figures~\ref{fig:sphere}(c,d) corresponds to an SPSW with $\ell=6$,
where $\ell$ is the degree of the spherical harmonic $Y_\ell^n$, and
that peak is common to the 13 $n$-values $-6\leq n\leq 6$. Note also
that the 14 eigenwavenumbers with $k_+\in[0,6]$, visible in
Figure~\ref{fig:sphere}(e), are true eigenwavenumbers in the reverse
plasmonic case, as confirmed to at least $13$ digits by comparison
with semi-analytic results. In contrast, ``HK 8-dens'' exhibits here,
in addition, around 25 false eigenwavenumbers and false
near-eigenwavenumbers. See Figure~\ref{fig:sphere}(f). Eleven of the
14 peaks in Figure~\ref{fig:sphere}(e) form a periodic pattern and
correspond to eigenwavenumbers with eigenfields that are SPSWs,
whereas the other three peaks correspond to eigenwavenumbers with
eigenfields that are not bound to the surface.

\subsection{Field images for the ``rotated starfish''}
\label{sec:starfish}

\begin{figure}[t!]
\centering
\includegraphics[height=50mm]{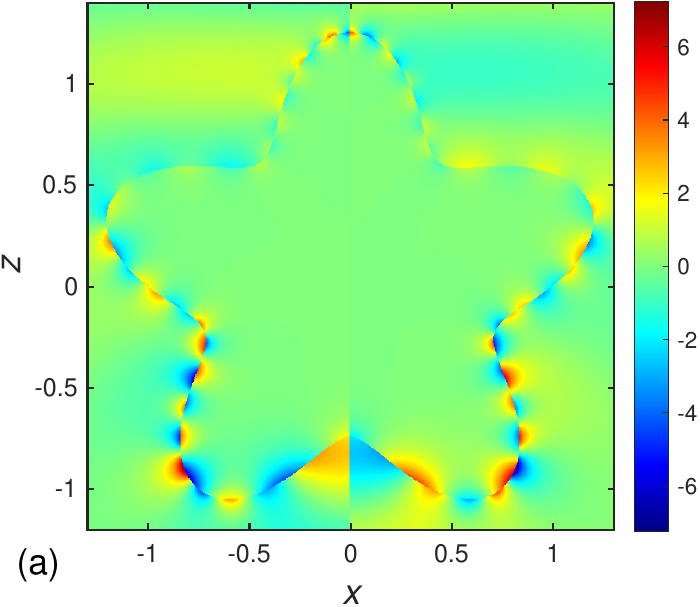}
\includegraphics[height=50mm]{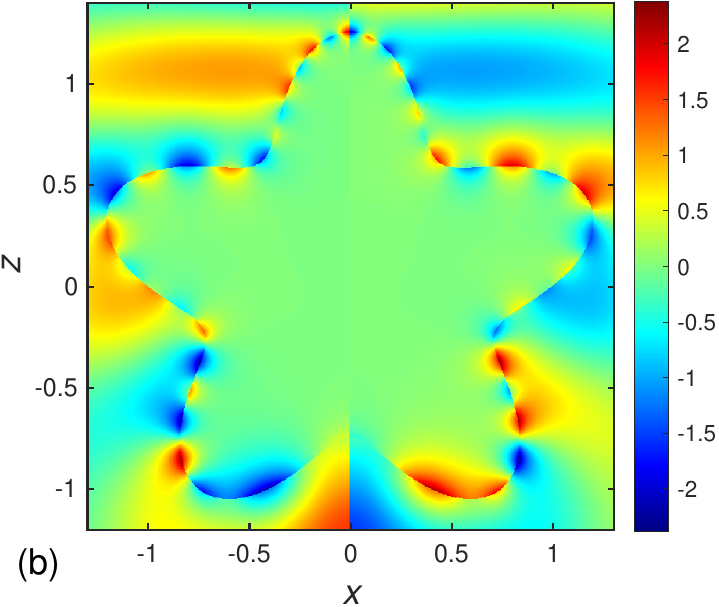}

\vspace{1mm}
\includegraphics[height=50mm]{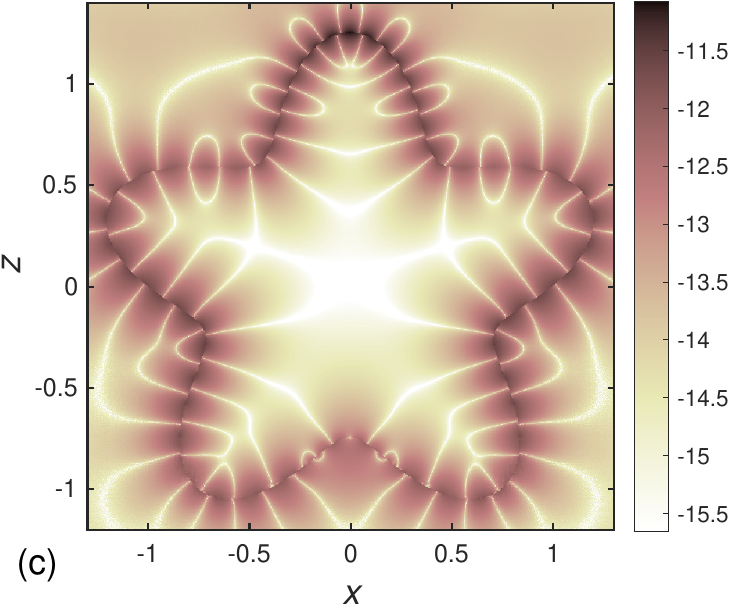}
\includegraphics[height=50mm]{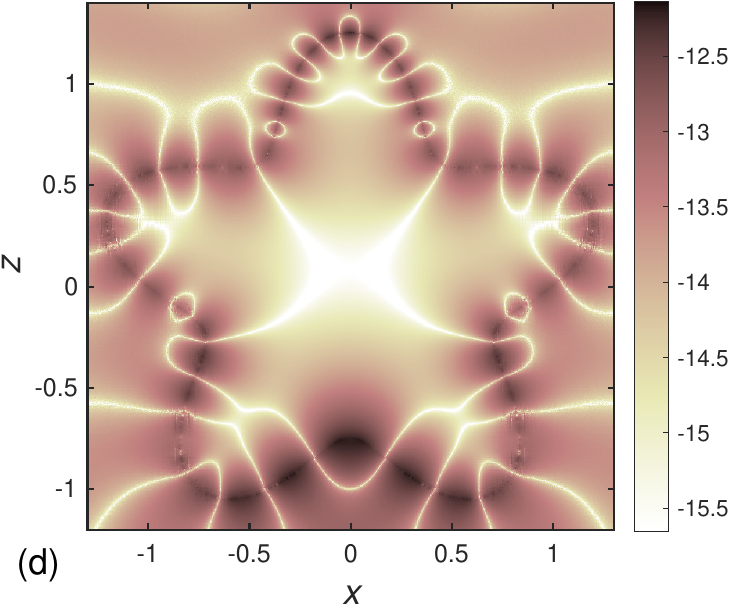}

\vspace{1mm}
\includegraphics[height=50mm]{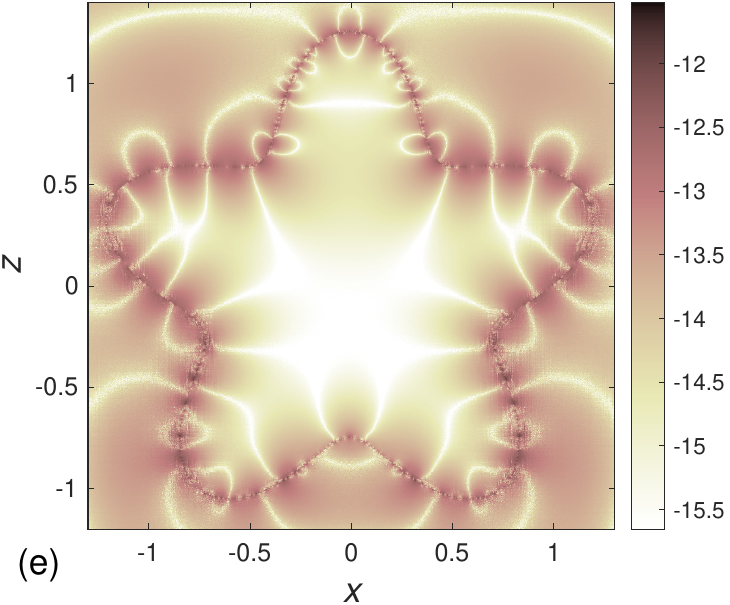}
\includegraphics[height=50mm]{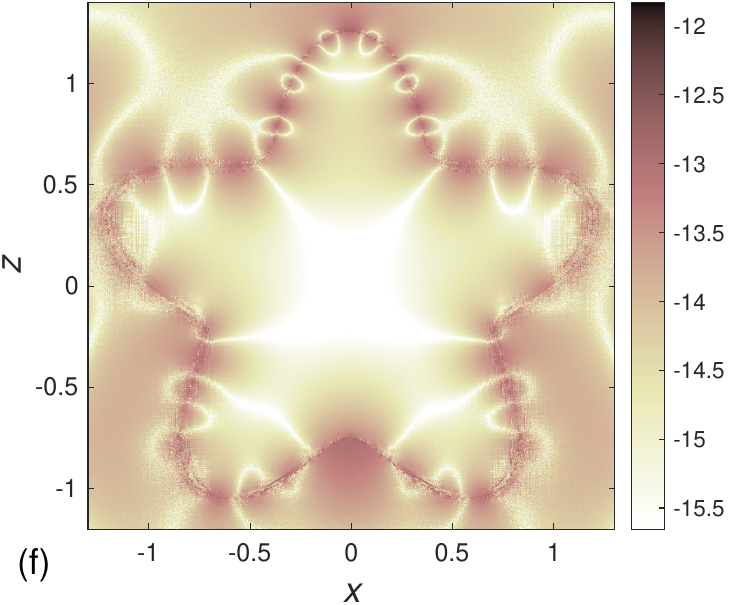}
\caption{\sf Field images for scattering from the ``rotated starfish'' 
  in the plasmonic case and with $\myvec E^{\rm in}(\myvec r)={\myvec
    x}e^{{\rm i}6z}$: (a,c,e) the scattered/transmitted $E_\rho(\myvec
  r,0)$-field; (b,d,f) the scattered/transmitted $H_\theta(\myvec
  r,0)$-field; (c,d) $\log_{10}$ of estimated absolute field error for
  ``HK 8-dens''; (e,f) the same for ``Dirac''.}
\label{fig:starfish}
\end{figure}

We compute images of scattered and transmitted fields $E^{\rm
  sc}_\rho(\myvec r,0)$, $H^{\rm sc}_\theta(\myvec r,0)$ and $E^{\rm
  tr}_\rho(\myvec r,0)$, $H^{\rm tr}_\theta(\myvec r,0)$. The surface
$\Gamma$ is that of the ``rotated starfish'', given
by~(\ref{eq:starfish}) with $\alpha=0.25$. A number of $768$
discretization points are again used for each integral operator on
$\gamma$ in~(\ref{eq:3DgenF}).

We first test the positive dielectric case ($\hat k=1.5$) with
$k_-=10$. This is a simple problem and both ``Dirac'' and ``HK
8-dens'' give fields with an estimated absolute precision of at least
twelve digits -- also close to $\Gamma$. We refrain from showing
images. The only noticeable difference between the methods is the
number of GMRES iterations needed for full convergence. The ``Dirac''
system converges to an estimated relative residual of $\epsilon_{\rm
  mach}$ in $111$ iterations while the ``HK 8-dens'' system is a bit
better and only needs $68$ iterations.

We next test the plasmonic case ($\hat k={\rm i}\sqrt{1.1838}$) with
$k_-=6$. Figure~\ref{fig:starfish} shows that the accuracy achieved by
``Dirac'' and ``HK 8-dens'' is similar also in this case, although the
spectral properties of ``Dirac'' are now much better than those of
``HK 8-dens''. The condition number of the system matrix in the
discretized ``Dirac'' system is $9.7\cdot 10^3$ while the
corresponding condition number for ``HK 8-dens'' is $1.2\cdot 10^5$
and this is reflected in the number of GMRES iterations needed for
full convergence. The ``Dirac'' system converges to an estimated
relative residual of $\epsilon_{\rm mach}$ in $170$ iterations
while the ``HK 8-dens'' system needs $623$ iterations.

The symmetry of $\myvec E^{\rm in}$ is such that only the two
azimuthal modes $n=-1$ and $n=1$ are present. The Fourier coefficients
of the surface densities of these modes, $h_{(1)}$ and $h_{(-1)}$
of~(\ref{eq:3DgenF}), are either identical or have opposite signs.
Animations based on~(\ref{eq:timedep}) for a sequence of $t$ reveal
that SPWs are propagating along $\Gamma$ in the images of
Figure~\ref{fig:starfish}(a,b). Their wavelength is roughly 15\%
shorter than the wavelength given by~(\ref{eq:SPW}).

\subsection{Convergence of field images} \label{sec:convergence}

\begin{figure}[t!]
\centering
\includegraphics[height=48mm]{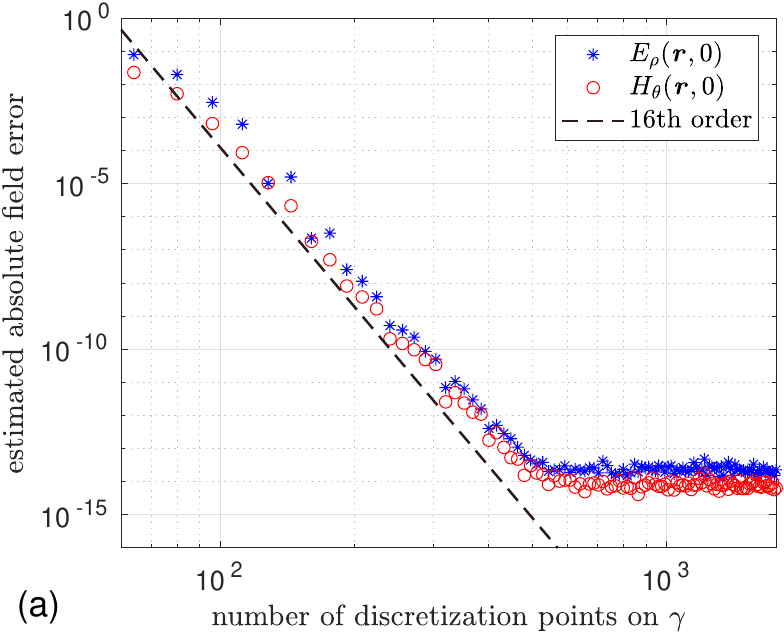}
\includegraphics[height=48mm]{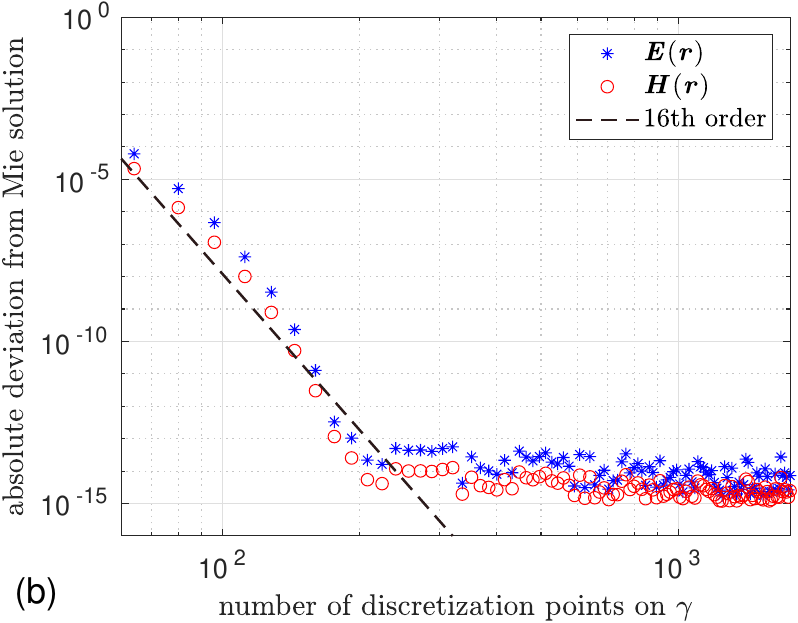}
\caption{\sf Convergence as a function of the number of
  discretization points on $\gamma$ of: (a) the scattered/transmitted
  $E_\rho(\myvec r,0)$-field and $H_\theta(\myvec r,0)$-field, shown
  in Figure~\ref{fig:starfish}(e,f); (b) the scattered/transmitted
  $\myvec E(\myvec r)$-field and $\myvec H(\myvec r)$-field resulting
  from the incident field~(\ref{eq:partial}) and $\Gamma$ being the
  unit sphere.} 
\label{fig:convor} 
\end{figure}

As pointed out by one of the referees, it is interesting to see how
the numerical error in field images, produced by our 16th order
Fourier--Nyström scheme of Section~\ref{sec:FouNysd} in conjunction
with ``Dirac'', evolves as the size of the discrete linear
systems~(\ref{eq:3DgenF}) increases. To this end,
Figure~\ref{fig:convor}(a) shows the average estimated absolute field
error in Figure~\ref{fig:starfish}(e,f) under mesh refinement. The
error is estimated using $90,\!000$ field points on a Cartesian grid
in the box ${\cal B}=\left\{-1.3\le x\le 1.3; -1.2\le z\le
  1.4\right\}$ and behaves very stably.

It is also of interest to make convergence studies on setups which
have semi-analytic solutions, and compare with these, even though such
solutions themselves may not be entirely free from numerical error. To
this end we let an incident field
\begin{equation}
\begin{gathered}
   \myvec E^{\rm in}(\myvec r)=\myvec G(\myvec r)
   +k_-^{-1}\nabla\times\myvec G(\myvec r)\,,
   \quad
   \myvec H^{\rm in}(\myvec r)=-{\rm i}\myvec E^{\rm in}(\myvec r)\,,\\
   \myvec G(\myvec r)=j_1(k_-|\myvec r|)\rho|\myvec
   r|^{-1}\myvec\theta\,, 
\end{gathered} 
\label{eq:partial} 
\end{equation}
be scattered from and transmitted into the unit ball. Here $j_1$ is
the first order spherical Bessel function of the first kind and we are
still in the plasmonic case with $\hat k={\rm i}\sqrt{1.1838}$ and
$k_-=6$. Figure~\ref{fig:convor}(b) shows the average absolute error,
obtained by comparison with the Mie solution, in the fields $\myvec
E(\myvec r)$ and $\myvec H(\myvec r)$ under mesh refinement. The error
is computed using $90,\!000$ field points on a Cartesian grid in the
box ${\cal B}=\left\{-2\le x \le 2; -2\le z\le 2\right\}$ and
normalized with the largest field amplitudes in ${\cal B}$. The
convergence in Figure~\ref{fig:convor}(b) is very similar to that in
Figure~\ref{fig:convor}(a).

\begin{figure}[t!]
\centering
\includegraphics[height=50mm]{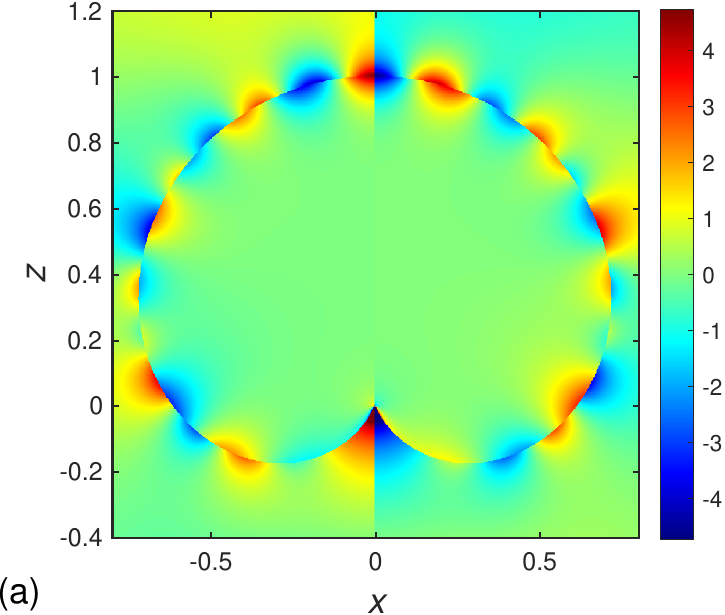}
\includegraphics[height=50mm]{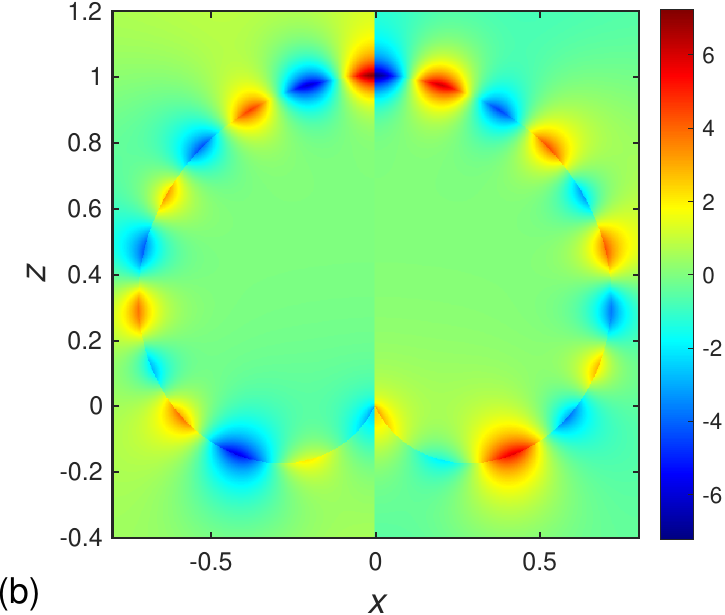}

\vspace{1mm}
\includegraphics[height=49mm]{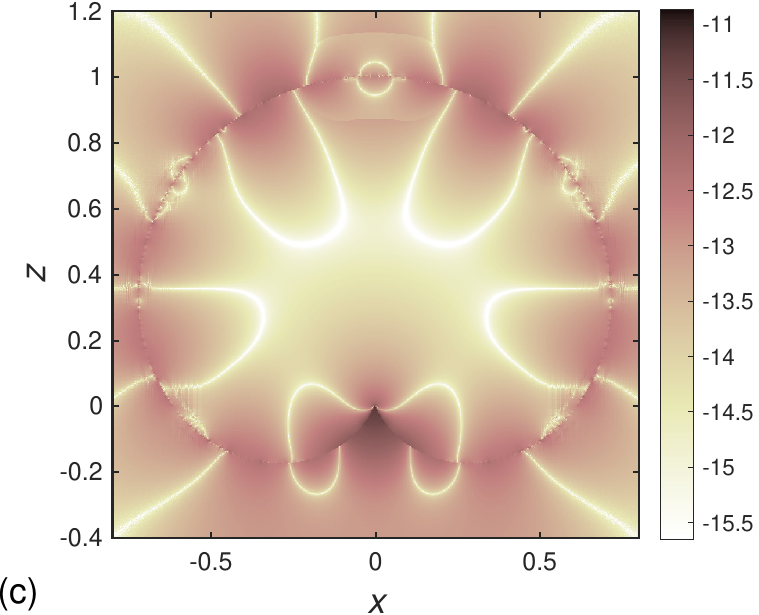}
\includegraphics[height=49mm]{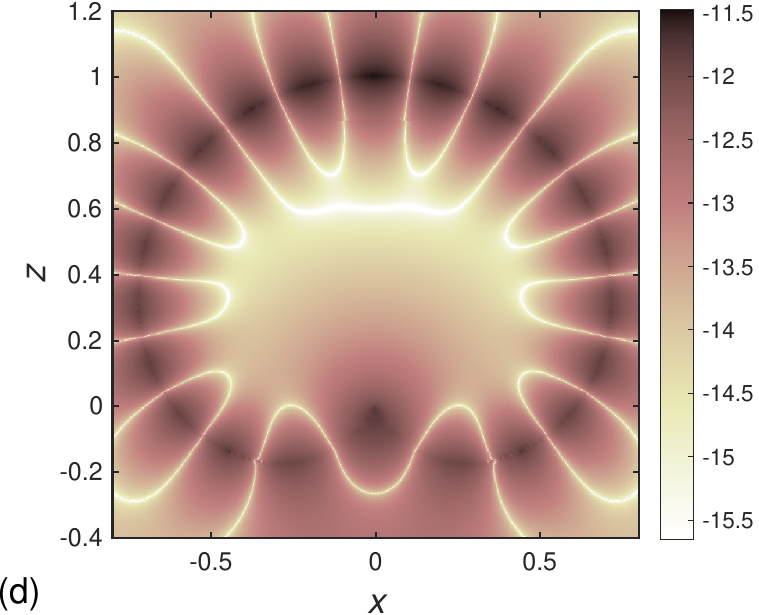}

\vspace{1mm}
\includegraphics[height=49mm]{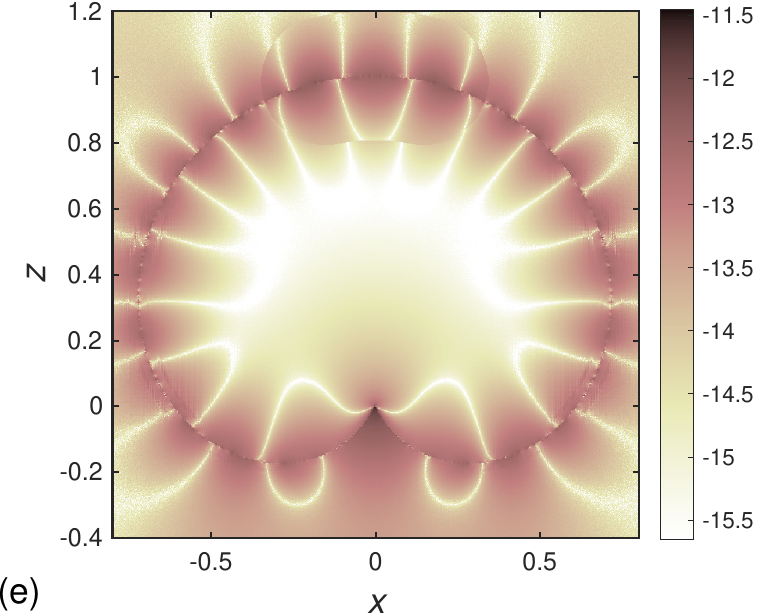}
\includegraphics[height=49mm]{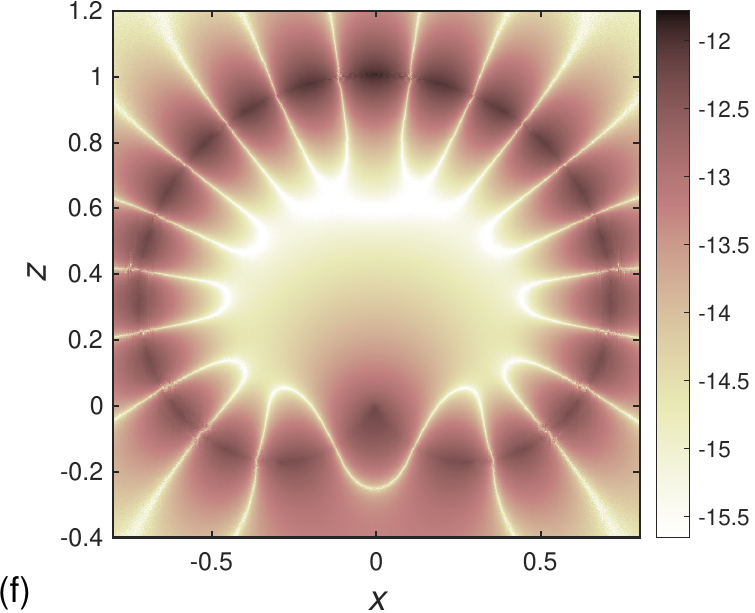}
\caption{\sf Field images for scattering from the ``tomato'' 
  in the plasmonic case and with $\myvec E^{\rm in}(\myvec r)={\myvec
    x}e^{{\rm i}5z}$: (a,c,e) the scattered/transmitted $E_\rho(\myvec
  r,0)$-field; (b,d,f) the scattered/transmitted $H_\theta(\myvec
  r,0)$-field; (c,d) $\log_{10}$ of estimated absolute field error for
  ``HK 8-dens''; (e,f) The same for ``Dirac''. The colorbar range in
  (a) is set to $[-4.55,4.55]$.}
\label{fig:tomato}
\end{figure}

\begin{figure}[t!]
\centering
\includegraphics[height=49mm]{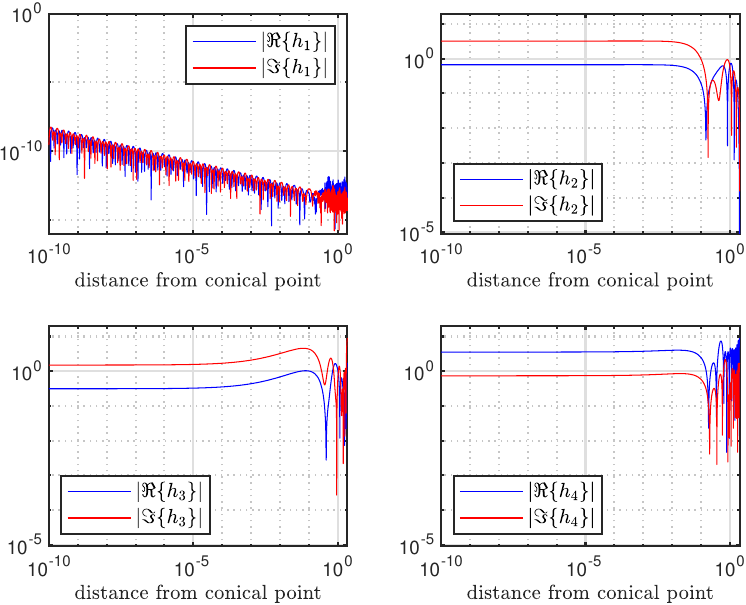}
\includegraphics[height=48mm]{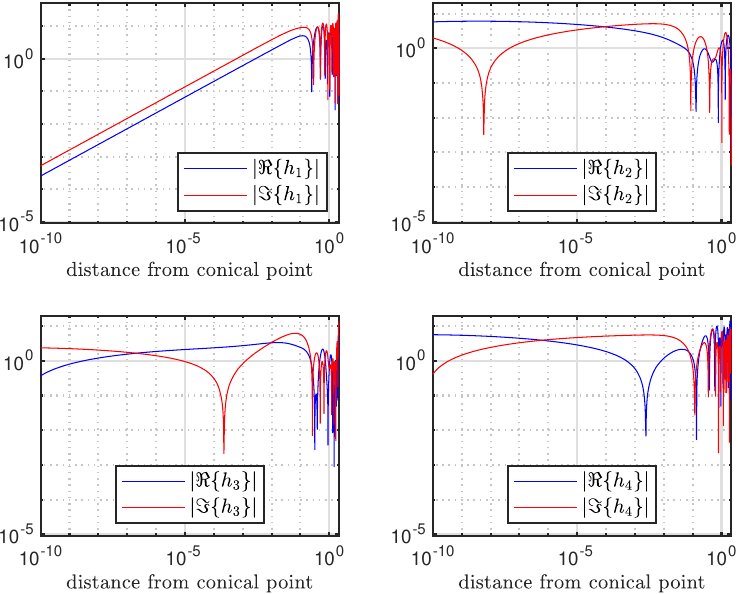}

\vspace{1mm}
\includegraphics[height=49mm]{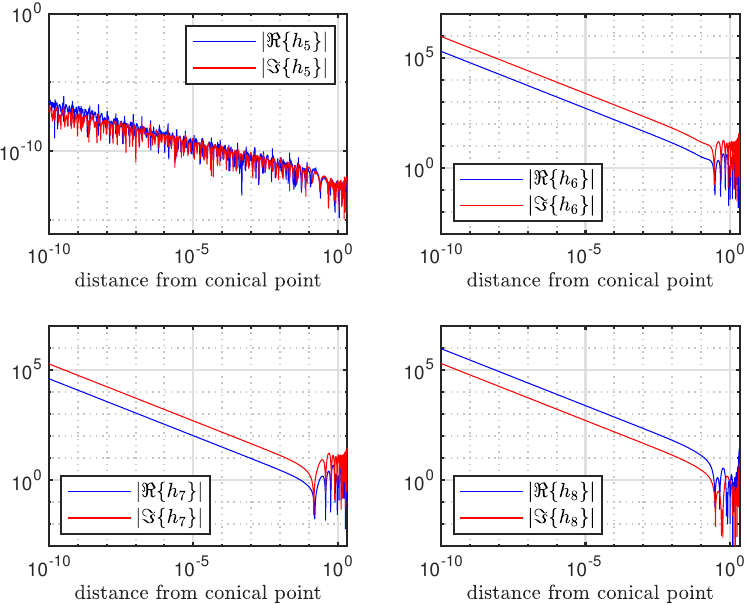}
\includegraphics[height=48mm]{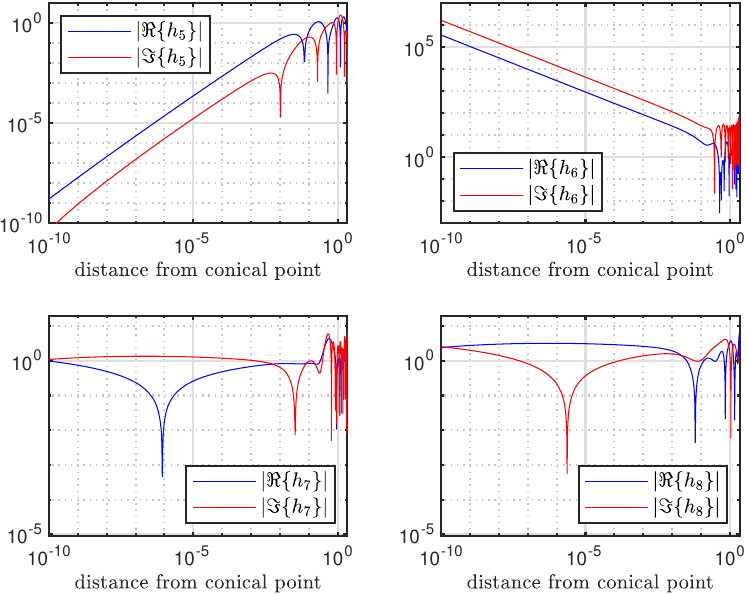}
\caption{\sf Asymptotics for the ``tomato'' in the plasmonic case. The
  densities $h_i$, $i=1,\ldots,8$, for azimuthal mode $n=1$, as
  functions of the arc length distance along $\gamma$ to the conical
  point at the origin. Columns 1-2: ``HK 8-dens''. Columns 3-4:
  ``Dirac''.}
\label{fig:tomatoasymp}
\end{figure}

\subsection{In a false essential spectrum for the ``tomato''}
\label{sec:tomato}

We repeat the experiments of Section~\ref{sec:starfish}, but now for
the ``tomato'', that is, with $\Gamma$ non-smooth and generated by
$\gamma$ of~(\ref{eq:tomato}) with $\alpha=31\pi/18$ as illustrated in
Figure~\ref{fig:amoeba0}(a). A number of $576$ discretization points
are used for each integral operator on $\gamma$ in~(\ref{eq:3DgenF}),
making the total system size $4608\times 4608$.

In the positive dielectric case, and with a larger wavenumber $k_-=18$
as to compensate for the ``tomato'' being smaller than the ``rotated
starfish'', the results for both methods are even (marginally) better
than in the previous example. The estimated pointwise precision in the
field images is between twelve and $13$ digits (no images shown). The
``Dirac'' system needs $96$ GMRES iterations for full convergence
while the ``HK 8-dens'' system needs $87$ iterations. We conclude
that, thanks to the RCIP method for dealing with the conical point,
the non-smooth ``tomato'' is as simple as the
smooth ``rotated starfish'' from a numerical point of view.

In the plasmonic case and with $k_-=5$ we observe some very
interesting features. The wavenumber ratio corresponds to that
$(1+\hat\epsilon)/(1-\hat\epsilon)$ is in $-\sigma_{\text{ess}}(K_{\rm
  d})$, but not in $\sigma_{\text{ess}}(K_{\rm d})$. Given the
discussion about essential spectra in Section~\ref{sec:ess}, it may
not come as a complete surprise that ``HK 8-dens'' exhibits false
essential spectrum in this case, even though correct limit solutions
can be computed, see the last paragraph of Section~\ref{sec:prop}.
``Dirac'', on the other hand, is free from this problem and correct
fields can be computed without a limit process.
Figure~\ref{fig:tomato} shows that ``Dirac'' clearly achieves better
field accuracy than ``HK 8-dens'' in this case. In terms of GMRES
convergence the difference is even greater: ``Dirac'' needs $88$
iterations while ``HK 8-dens'' needs $326$ iterations. There are SPWs
propagating along $\Gamma$ with a wavelength that is roughly 20\%
shorter than the wavelength given by~(\ref{eq:SPW}).

It is also enlightening to inspect the asymptotics of $h$ close to the
conical tip of the ``tomato'' in the plasmonic case. Thanks to the
symmetry of $\myvec E^{\rm in}$, as discussed in the last paragraph of
Section~\ref{sec:starfish}, it is enough to study the eight densities
(Fourier coefficients) contained in the modal solution $h_{(1)}$.
Figure~\ref{fig:tomatoasymp} shows these eight densities both for
``Dirac'' and ``HK 8-dens''. The individual densities are denoted
$h_i$, $i=1,\ldots,8$, with the azimuthal index omitted. Note that the
densities $h_1$ and $h_5$ of ``HK 8-dens'' are approximately zero, as
they should be according to Section~\ref{sec:HK8dens}. 

\begin{figure}[t!]
\centering
\includegraphics[height=50mm]{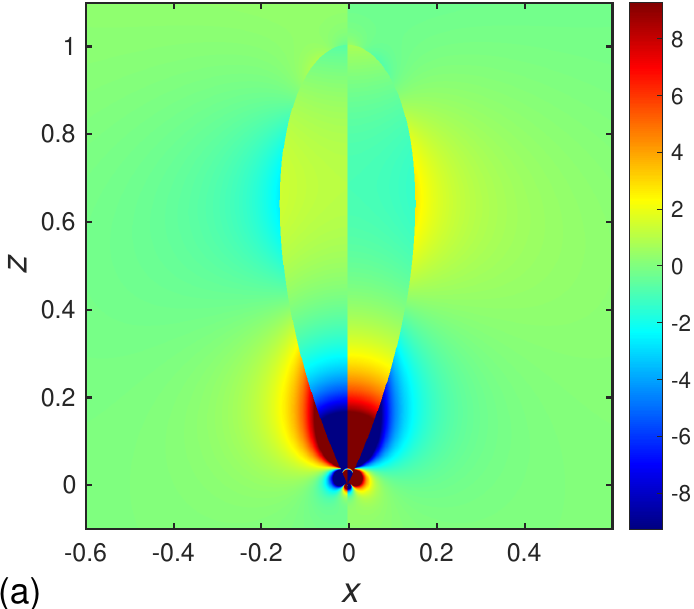}
\hspace{2mm}
\includegraphics[height=50mm]{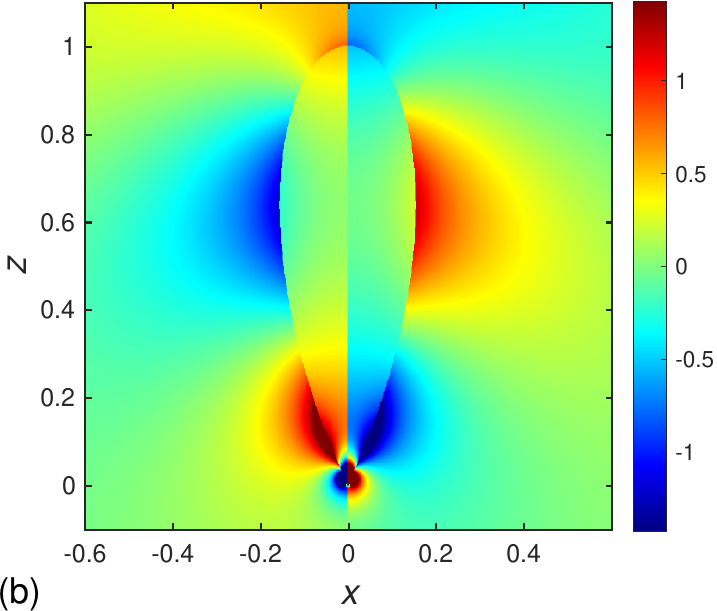}

\vspace{1mm}
\includegraphics[height=50mm]{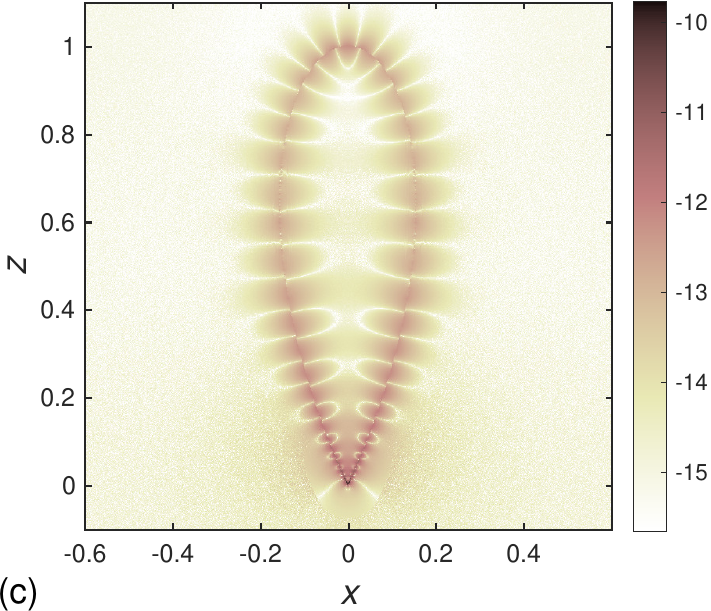}
\hspace{2mm}
\includegraphics[height=50mm]{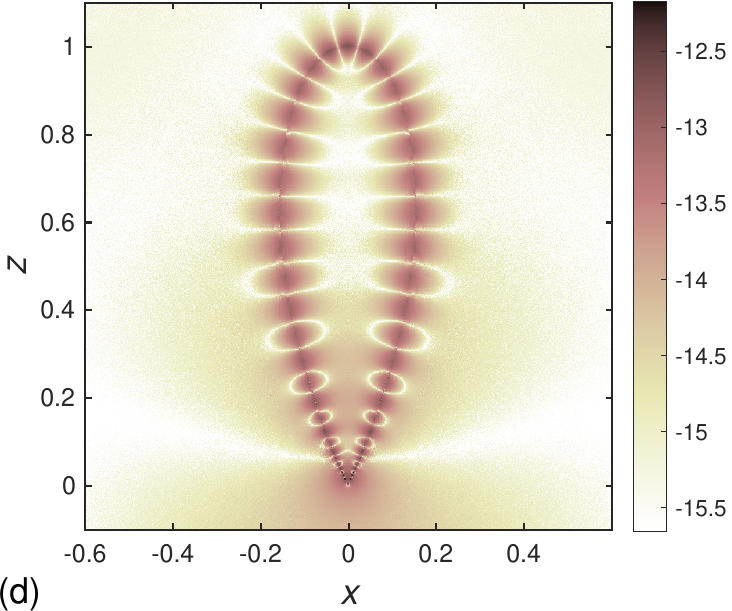}
\caption{\sf Field images for scattering from the ``drop with a sharp tip'' 
  in the plasmonic case and with $\myvec E^{\rm in}(\myvec r)={\myvec
    x}e^{{\rm i}5z}$: (a,c) the scattered/transmitted
  $E^\downarrow_\rho(\myvec r,0)$-field; (b,d) the
  scattered/transmitted $H^\downarrow_\theta(\myvec r,0)$-field; (c,d)
  $\log_{10}$ of estimated absolute field error for ``Dirac''. The
  colorbar range in (a) is set to $[-9.27,9.27]$ and in (b) set to
  $[-1.43,1.43]$.}
\label{fig:drop}
\end{figure}

The strongest singularity observed in Figure~\ref{fig:tomatoasymp} is
\begin{equation}
h_i\propto s^{-0.51712968815959}\,,
\label{eq:singtom}
\end{equation}
where $s$ is the arc length distance along $\gamma$ to the conical
point at the origin. The exponents in \eqref{eq:singtom} and
\eqref{eq:singdrp} are estimated using the automated eigenvalue method
of \cite[Section 14]{Helsing18} and all displayed digits are believed
to be correct. For ``HK 8-dens'', the singularity
of~(\ref{eq:singtom}) is observed for the densities $h_6$, $h_7$, and
$h_8$, corresponding to $\varrho_{\rm E}$, $M_\theta$, and $M_\tau$
according to~(\ref{eq:physint}). For ``Dirac'', the singularity
of~(\ref{eq:singtom}) is observed only for the density $h_6$. Note
that each of these functions is more singular than $H^{1/2}(\Gamma)$,
as allowed by the function space $\mH_3$.

\begin{figure}[t!]
\centering
\includegraphics[height=49mm]{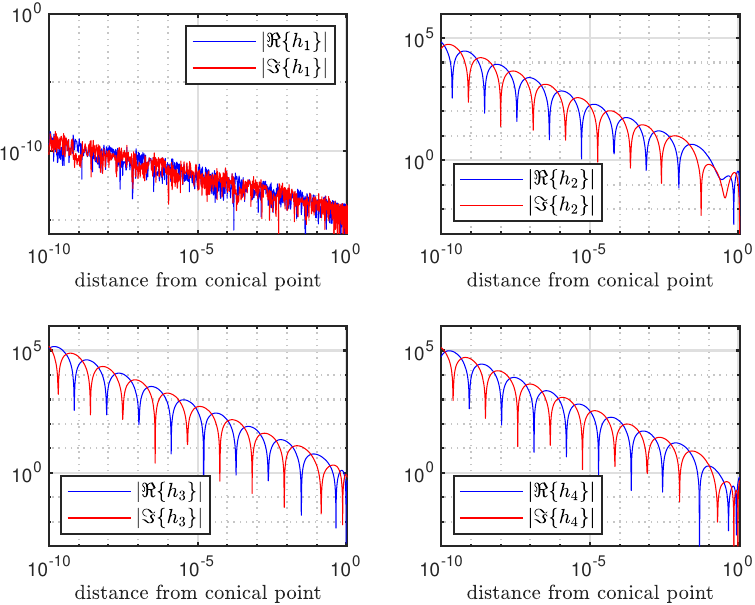}
\includegraphics[height=48mm]{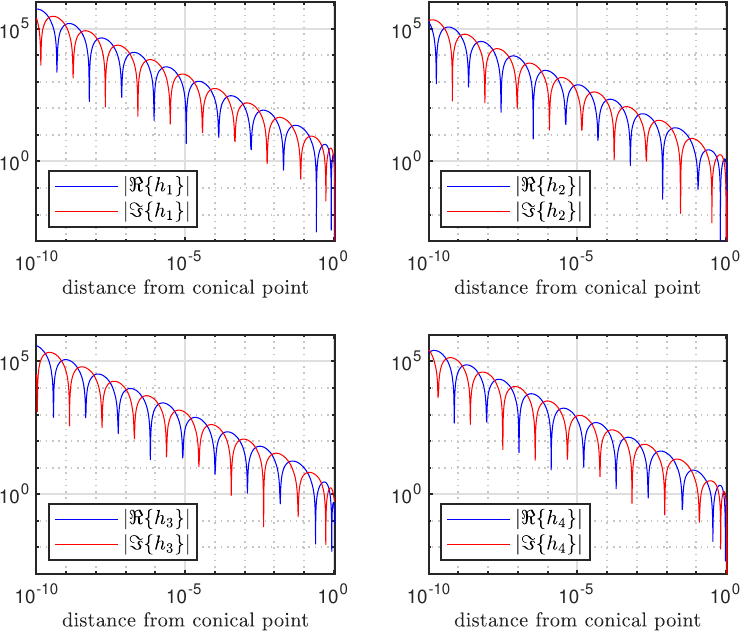}

\vspace{1mm}
\includegraphics[height=48mm]{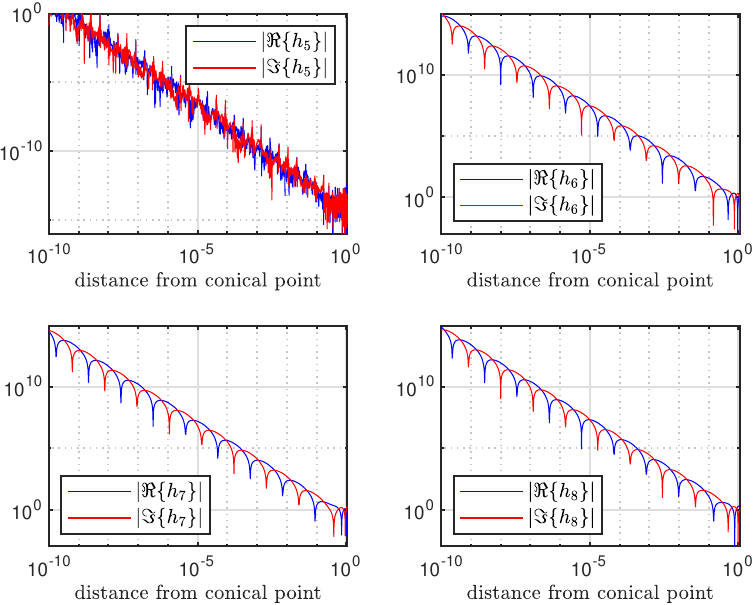}
\includegraphics[height=47mm]{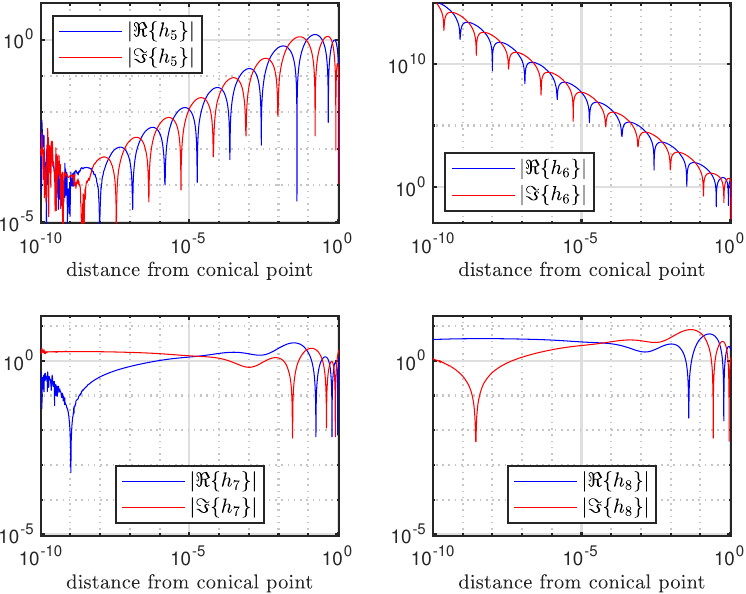}
\caption{\sf Asymptotics for the ``drop with a sharp tip'' in the 
  plasmonic case. Densities displayed as in
  Figure~\ref{fig:tomatoasymp}. Column 1-2: ``HK 8-dens''. Column 3-4:
  ``Dirac''.}
\label{fig:dropasymp}
\end{figure}

\subsection{In the essential spectrum for a ``drop with a sharp tip''}
\label{sec:drop}

We repeat some of the experiments of Section~\ref{sec:tomato}, but
shrink the opening angle of the conical point to $\alpha=5\pi/18$, so
that the shape of $\Gamma$ now resembles that of a ``drop with a sharp
tip''. A number of $576$ discretization points are again used for each
integral operator on $\gamma$ in~(\ref{eq:3DgenF}).

Field images for the plasmonic case along with error estimates for
``Dirac'' are shown in Figure~\ref{fig:drop}. No SPWs are excited
along $\Gamma$. A number of $83$ iterations were needed for full GMRES
convergence. The corresponding number for ``HK 8-dens'' is $314$
iterations. Note that $(1+\hat\epsilon)/(1-\hat\epsilon)$ is now in
the essential spectrum of $K_{\rm d}$ and that $E^\downarrow_\rho$ and
$H^\downarrow_\theta$ are oscillatory and unbounded at the origin.
The colorbar ranges in Figure~\ref{fig:drop}(a,b) are therefore
restricted to the most extreme field values away from the origin. The
estimated pointwise absolute error close to the origin is, of course,
now larger than than in previous examples with everywhere bounded
fields.

Figure~\ref{fig:dropasymp} is analogous to
Figure~\ref{fig:tomatoasymp} and exhibits the same general features.
The strongest singularity observed is now
\begin{equation}
h_i\propto s^{-1.5+1.25455347163480{\rm i}}\,,
\label{eq:singdrp}
\end{equation}
clearly visible for $h_6$ of ``Dirac'' and for $h_6$, $h_7$, and $h_8$
of ``HK 8-dens''.
As we are now in the essential spectrum, we do not expect the densities
to belong to $\mH_3$, and indeed the above mentioned densities just
fail to belong to $H^{-1/2}(\Gamma)$. Moreover,
the densities $h_3$ and $h_4$, and for ``Dirac'' also $h_1$, just fail 
 to belong to $H^{1/2}(\Gamma)$. This second strongest singularity observed
is $s^{-0.50000+1.25455{\rm i}}$.
 Again $h_1$ and $h_5$ are zero for ``HK 8-dens'', modulo rounding
 errors.

\subsection{Static plasmons}
\label{sec:staticplots}

\begin{figure}[t!]
\centering
\includegraphics[height=45mm]{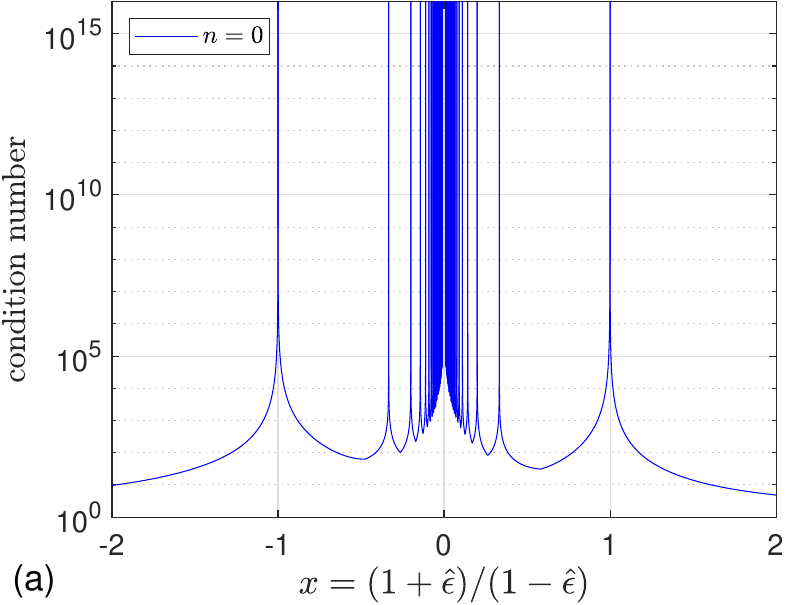}
\includegraphics[height=45mm]{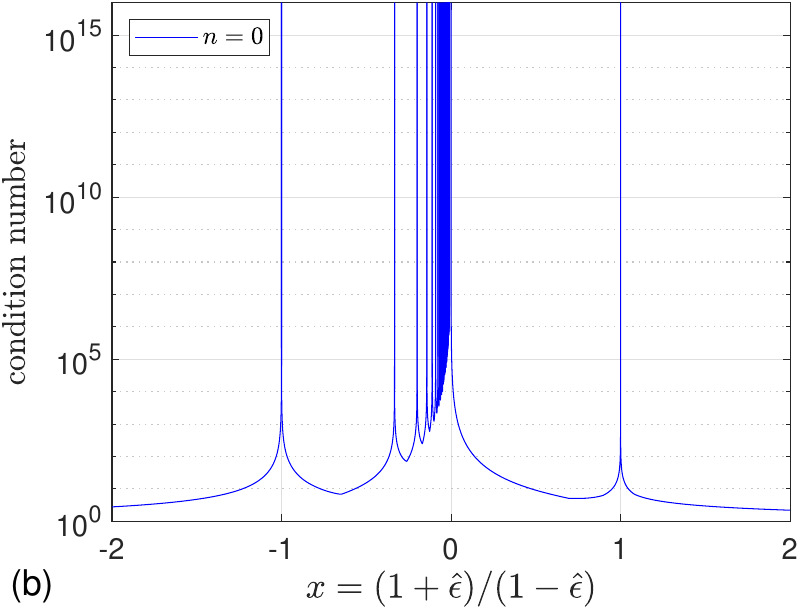}

\vspace{1mm}
\includegraphics[height=45mm]{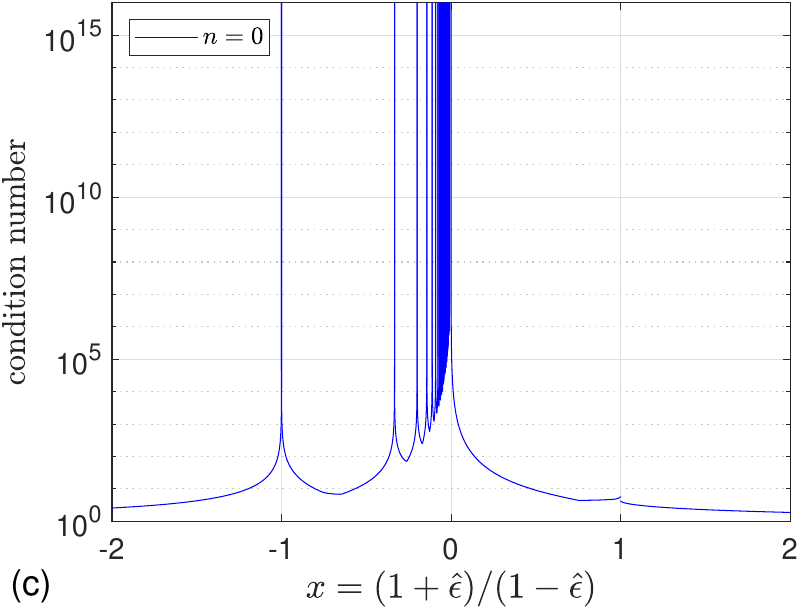}
\includegraphics[height=45mm]{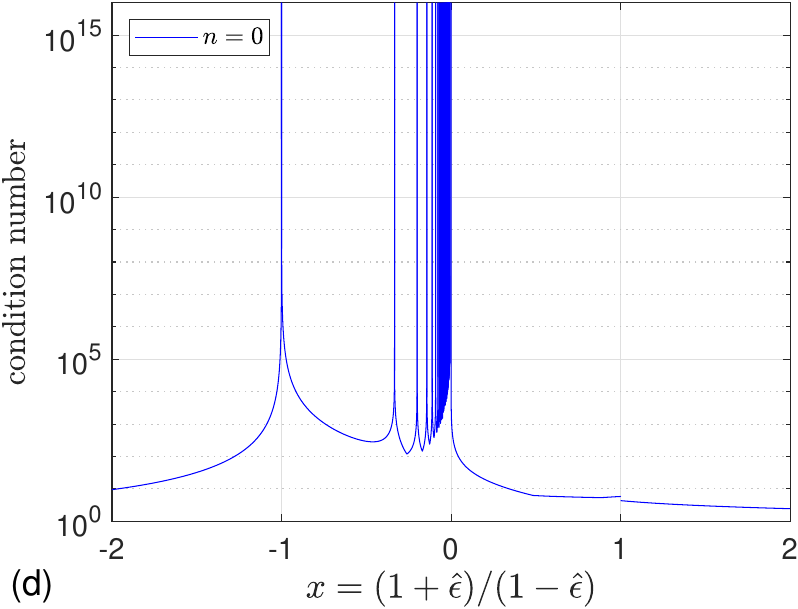}
\caption{\sf The unit sphere. Condition numbers of system matrices of 
  the discretized modal integral equations~(\ref{eq:3DgenF}) with
  azimuthal index $n=0$ in the quasi-static limit $k_+=k_-=0$. The
  plasmonic case is $x\in(-1,1)$ and the positive dielectric case is
  $x\in (-2,-1)\cup (1,2)$. (a) ``HK 8-dens''; (b) ``Dirac''; (c) The
  ``Dirac'' (5:8,5:8) diagonal block; (d) The ``Dirac'' version
  \eqref{eq:pecdir}.}
\label{fig:quasistat}
\end{figure}

Similar to Section~\ref{sec:sphere} we plot condition numbers of our
IERs for the unit sphere to detect possible false eigenwavenumbers,
but also to get a notion of how well-conditioned the two IERs under
study are. This time we consider the quasi-static limit $k_\pm\to 0$
of the IERs as in Section~\ref{sec:static}, and plot the condition
numbers as functions of $x=(1+\hat\epsilon)/(1-\hat\epsilon)$ in
Figure~\ref{fig:quasistat}. With a slight abuse of notation, we keep
speaking of (true/false) eigenwavenumbers.

Here $x<-1$ corresponds to the positive dielectric case $\hat k>1$,
$x=-1$ corresponds to an exterior Dirichlet problem, $-1<x<0$
corresponds to the plasmonic case $\hat k/{\rm i}>1$, $x=0$
corresponds to the essential spectrum, $0<x<1$ corresponds to the
plasmonic case $0<\hat k/{\rm i}<1$, $x=1$ corresponds to an exterior
Neumann problem, and $x>1$ corresponds to the positive dielectric case
$0<\hat k<1$.
 
As discussed in Section~\ref{sec:static}, the condition number of the
(5:8,5:8) diagonal block for ``Dirac'', shown in
Figure~\ref{fig:quasistat}(c), follows closely that of $K_{\rm d}$ and
the peaks correspond to the static plasmons discussed in
Section~\ref{sec:plasmons}. The peak at $x=-1/3$ has a bright plasmon
whereas all others have dark plasmons. ``HK 8-dens'' shows a symmetric
spectrum with an infinite number of false eigen\-wavenumbers for
$0<x<1$. The full ``Dirac'' shows only two false eigen\-wavenumbers in
the limits $x=\pm 1$. As shown in Figure~\ref{fig:quasistat}(d), the
peak at $x=1$ can be removed by using the version of ``Dirac''
specified by \eqref{eq:pecdir}. Note that ``Dirac'' specified by
\eqref{eq:3DPN}, is better conditioned near $x=-1$, where false
eigenwavenumbers occur both for ``Dirac'' and ``HK 8-dens''. For
``Dirac'' this false eigenwavenumber corresponds to the monopole field
$\myvec E=\myvec r/|\myvec r|^3$, $\myvec H=\myvec 0$ for $|\myvec
r|>1$, and $\myvec E=\myvec H=\myvec 0$ for $|\myvec r|<1$.

\section{Conclusions and discussion}
\label{sec:conclusion}

Our numerical results show that ``Dirac'' wins over ``HK 8-dens'' in
almost all tests. ``Dirac'' does not have any false
near-eigenwavenumbers for any passive materials, whereas ``HK 8-dens''
exhibits such in the plasmonic case, and even false eigenwavenumbers
in the reverse plasmonic case. We have seen that ``HK 8-dens''
exhibits false essential spectrum on domains with corners, and false
eigenwavenumbers in the quasi-static limit. ``HK 8-dens'' also
requires more than three times as many GMRES iterations than ``Dirac''
in the plasmonic cases.

Since ``Dirac'' clearly is a competitive IER of the MTP($k_-,k_+$), we
end by comparing it to other available IERs. We limit our discussion
to non-magnetic materials, although ``Dirac'' was formulated for
general materials in \cite{HelsRose20}, which we plan to cover in a
forthcoming publication. The aspects of the IERs that we base our
discussion on are the following.
\begin{itemize}
\item[{\rm (A)}] Has it been used for Lipschitz regular/piecewise
  smooth surfaces $\Gamma$?
\item[{\rm (B)}] What types of operators does it employ?
\item[{\rm (C)}] How fast does it compute all the scattered and
  transmitted fields?
\item[{\rm (D)}] How does it behave in the quasi-static limit?
\item[{\rm (E)}] How stable is it? Does it have false
  eigenwavenumbers, false near-eigenwavenumbers, or false essential
  spectrum?
\end{itemize} 
We now make a comparison with the ``Debye'' formalism in
\cite{EpsGreNei13,EpsGreNei19} and the ``DFIE'' formalism in
\cite{VicGreFer18} which, according to
\cite[Conclusions]{VicGreFer18}, are the current leading IERs when it
comes to well-posedness for a wide range of passive materials. Using
formalism from \cite{HelsRose20,RosenGMA19}, ``Dirac'' uses Cauchy
type integral representations for the Dirac equation $\dirac F= {\rm
  i}kF$ for the electromagnetic multivector field $F$, whereas
``Debye'' employs a Dirac multivector potential $F= \dirac G+{\rm
  i}kG$, leading to a component-wise Helmholtz equation $\Delta G+ k^2
G=0$, for which classical layer potentials apply. The electric and
magnetic Gauss equations translate to certain conditions on these
boundary layers, which via Hodge decompositions of tangential vector
fields allows one to eliminate all but two scalar boundary densities
for each domain. In the scattering situation of the present paper,
this yields a ``Debye'' IER with four scalar densities. It is known
that (D) it does not have dense-mesh low-frequency breakdown, and no
topological low-frequency breakdown if complemented by extra
equations. A drawback of ``Debye'' is that implementing the Hodge
projections requires the numerical solution of a second order surface
Laplace equation on $\Gamma$. An implementation of ``Debye'' for
scattering by perfect conductors and piecewise smooth $\Gamma$ is
in~\cite{ChernoBoag13}, but (A) it has not been used for the
MTP($k_-,k_+$) on piecewise smooth $\Gamma$. According to
\cite[Equation (2.19)]{EpsGreNei19}, ``Debye'' also (B) uses
compositions of two integral operators. Concerning (E), it is stated
in \cite[Conclusions]{EpsGreNei19} that ``Debye'' is invertible for
all passive materials. However, this is impossible for any IER, which
is equivalent to MTP($k_-$, $k_+$), due to the true eigenwavenumbers
shown in Figure~\ref{fig:sphere}(e). Moreover, the main uniqueness
result for ``Debye'', \cite[Theorem~3.2]{EpsGreNei13}, assumes that
$\Re{\rm e}\left\{\epsilon_+\right\}>0$ and relies on \cite[Theorem
69]{Muller69}, which assumes permittivities in the first quadrant
\cite[p.~261]{Muller69}. These assumptions do not cover all passive
non-magnetic materials, which correspond to the square $[0,\pi/2]^2$
in Figure~\ref{fig:amoeba0}(b). It was recently shown in
\cite{EpsteinRachh21} though, that these uniqueness results can be
sharpened to cover $(\arg(k_-),\arg(k_+))=(0,\pi/2)$ in particular.
However, numerical experiments revealing possible false
near-eigenwavenumbers as in Figure~\ref{fig:sphere}(d) are not
available.

``DFIE'' is formulated in \cite{VicGreFer18}. Like ``Debye'', it
employs potentials to reduce Maxwell's equations to vector Helmholtz
equations. With ``DFIE'', the fields $\myvec E$ and $\myvec H$ are
calculated independently of each other, by solving an IER with six
scalar densities for each field. Again it is known for ``DFIE'' that
(D) it does not suffer from low-frequency breakdown and (E) it does
not have false eigenwavenumbers for $k_->0$ and $0\le \arg (\hat
\epsilon)<\pi$, but it has false eigenwavenumbers at $(\arg(k_-),
\arg(k_+))=(0,\pi/2)$ as shown in \cite[Section~11.3.2]{HelsKarl19}
(at least for a 2D version of ``DFIE''). The theory in
\cite{VicGreFer18} is presented for H\"older boundary function spaces,
and it is assumed that $\Gamma$ is $C^2$ regular. However, by
straightforward adaption to fractional Sobolev spaces and Lipschitz
regular $\Gamma$ as in \cite{HelsRose20}, it appears that (A) ``DFIE''
is applicable to non-smooth scattering. Inspecting the integral
operators that ``DFIE'' employs in \cite[Appendix D]{VicGreFer18}, we
see that (B) it uses single and double layer type boundary operators
like ``Dirac'', with one exception. The exception is $K_{31}$ in
\cite[Equation (89)]{VicGreFer18}, which is a compact difference of
hypersingular operators, and somewhat more numerically challenging to
implement~\cite{LaiOneil19}. To compare (C) how fast ``DFIE'' and
``Dirac'' are, since the operators employed are roughly the same and
having a good solver in mind which is linear in the number of non-zero
operator blocks, we count these. From \cite[Equation
(87)]{VicGreFer18} we have $2(36-4)= 64$ non-zero operator blocks for
``DFIE'', keeping in mind that we need to solve two $6\times 6$
systems for obtaining both $\myvec E$ and $\myvec H$. From
\eqref{eq:Ek3D} we see that ``Dirac'' uses $64-14=50$ non-zero
operator blocks. For field evaluations, \cite[Equations (36) and
(38)]{VicGreFer18} show that ``DFIE'' requires the evaluation of three
double layer and three single layer potentials, for each component of
the fields. For ``Dirac'', using the projected densities
\eqref{eq:tweakeddens}, which are fast to compute on $\Gamma$, each
field component requires in general the evaluation of three double
layer and two single layer potentials.

Comparing with ``Dirac'', this has the advantage of (A) applying to
any Lipschitz regular $\Gamma$ and (B) only using bounded integral
operators of single and double layer type with explicit kernels. We
remark that although \eqref{eq:3Dgen} seems to be a Fredholm equation
of the second kind, the operator $G$ is not compact, even on smooth
$\Gamma$. However $G^4$ is compact, which explains why ``Dirac'' works
well in an iterative solver. Moreover, ``Dirac'' does not (E) have any
false eigenwavenumbers, false near-eigenwavenumbers or false essential
spectrum, (D) not even in the quasi-static limit, except at $x=\pm 1$.
These quasi-static endpoints are related to eddy current scattering,
and is the topic of a forthcoming paper by the authors. A reason why
``Dirac'' is so successful in avoiding false spectra, is that the
choices of parameters make its invertibility depend only on the (6,6)
diagonal block involving $K_{\rm d}$. Thus the false spectrum of
$K_{\rm m}$, of which only one Hodge component is needed for the
MTP($k_-,k_+$), is avoided.

In~\cite[Important Remark p.~123]{Ammari16} it is stated that
$\sigma(K_{\rm m})\setminus\sigma(K_{\rm d})$ contributes to
``higher-order resonances'' in the near quasi-static limit. However,
plots of the condition numbers for ``Dirac'' on the sphere near the
quasi-static limit (not shown) are very similar to
Figure~\ref{fig:quasistat}(b). In particular, there are no visible
traces of $\sigma(K_{\rm m})$ in terms of near-eigenwavenumbers for
$0<x<1$.

\section*{Acknowledgement}

\noindent
This work was supported by the Swedish Research Council under contract
2015-03780.

%\clearpage\newpage
%\vspace{9mm}
%\vspace{23mm}
%\vspace{18mm}

%\newpage
%\allowdisplaybreaks
%\appendix
%\centerline{*** {\LARGE{\bf Appendix}} ***}
%\section{The entries of $E_k$}
%\label{app:A}

\vspace{15mm}
%\newpage
\appendix
\centerline{*** {\LARGE{\bf Appendix}} ***}
\section{The entries of $E_k$}
\label{app:A}

The entries of the matrix $E_k$ of~(\ref{eq:Ek3D}) involve two
families of integral operators denoted $S_k$ and $K_k$. A given member
in an operator family, $S_k^\alpha$ or $K_k^\alpha$, is defined by its
superscript $\alpha$, which can be a constant, a unit vector, a scalar
product of unit vectors, or a cross product of unit vectors.
Specifically we have for $S_k^1$ acting on a general density $g$
\begin{equation}
S_k^{1}g(\myvec r)={\rm i}k
\int_\Gamma\Phi_k(\myvec r,\myvec r')g(\myvec r')\,{\rm d}\Gamma'\,,
\quad \myvec r\in\Gamma\,,
\end{equation}
for $S_k^{\myvec u\cdot \myvec v'}$, where $\myvec u$ and $\myvec v$
are unit vectors,
\begin{equation}
S_k^{\myvec u\cdot \myvec v'}g(\myvec r)={\rm i}k
\int_\Gamma\myvec u(\myvec r)\cdot\myvec v(\myvec r')
\Phi_k(\myvec r,\myvec r')g(\myvec r')\,{\rm d}\Gamma'\,,
\quad \myvec r\in\Gamma\,,
\end{equation}
and for $K_k^{ \myvec u}$ and $K_k^{\myvec u\times\myvec v'}$
\begin{align}
K_k^{\myvec u}g(\myvec r)&=
\text{\rm p.v.\!}
\int_\Gamma\myvec u(\myvec r)\cdot\nabla'\Phi_k(\myvec r,\myvec r') 
g(\myvec r')\,{\rm d}\Gamma'\,,
\quad\myvec r\in\Gamma\,,\\
K_k^{\myvec u\times\myvec v'}g(\myvec r)&=
\text{\rm p.v.\!}
\int_\Gamma ( \myvec u(\myvec r)\times\myvec v(\myvec r') ) \cdot\nabla' 
\Phi_k(\myvec r,\myvec r')g(\myvec r')\,{\rm d}\Gamma'\,,
\quad\myvec r\in\Gamma\,.
\end{align}

When coding, particularly when $\Gamma$ is axially symmetric and when
azimuthal Fourier transforms are to be implemented, it is helpful to
have $S_k^\alpha$ and $K_k^\alpha$ in the form
\begin{align}
S_k^\alpha g({\myvec r})&=
{\rm i}k\int_\Gamma \frac{s_\alpha(\myvec r,\myvec r')}
                 {2\pi\vert\myvec r-\myvec r'\vert}
e^{{\rm i}k\vert\myvec r-\myvec r'\vert}
g({\myvec r'})\,{\rm d}\Gamma'\,,
\quad \myvec r\in\Gamma\,,
\label{eq:Sdef}\\
K_k^\alpha g({\myvec r})&=
\text{\rm p.v.\!}
\int_\Gamma \frac{d_\alpha(\myvec r,\myvec r')}
                 {2\pi\vert\myvec r-\myvec r'\vert^3}
(1-{\rm i}k\vert\myvec r-\myvec r'\vert)
e^{{\rm i}k\vert\myvec r-\myvec r'\vert}
g({\myvec r'})\,{\rm d}\Gamma'\,,
\quad \myvec r\in\Gamma\,,
\label{eq:Kdef}
\end{align}
where $s_\alpha(\myvec r,\myvec r')$ and $d_\alpha(\myvec r,\myvec
r')$ are static kernel factors expressed in terms of quantities
introduced in~(\ref{eq:2D1}--\ref{eq:2D3}) and the azimuthal angle
$\theta$.

Here follow $s_\alpha(\myvec r,\myvec r')$ for the ten operators
$S_k^\alpha$ of $E_k$
\begin{align}
s_1(\myvec r,\myvec r')&=1\,,\\
s_{\myvec\nu\cdot\myvec\nu'}(\myvec r,\myvec r')&=
\nu_\rho\nu_\rho'\cos(\theta-\theta')+\nu_z\nu_z'\,,\\
s_{\myvec\nu\cdot\myvec\theta'}(\myvec r,\myvec r')&=
\nu_\rho\sin(\theta-\theta')\,,\\
s_{\myvec\nu\cdot\myvec\tau'}(\myvec r,\myvec r')&=
\nu_\rho\nu_z'\cos(\theta-\theta')-\nu_z\nu_\rho'\,,\\
s_{\myvec\tau\cdot\myvec\nu'}(\myvec r,\myvec r')&=
\nu_z\nu_\rho'\cos(\theta-\theta')-\nu_\rho\nu_z'\,,\\
s_{\myvec\tau\cdot\myvec\theta'}(\myvec r,\myvec r')&=
\nu_z\sin(\theta-\theta')\,,\\
s_{\myvec\tau\cdot\myvec\tau'}(\myvec r,\myvec r')&=
\nu_z\nu_z'\cos(\theta-\theta')+\nu_\rho\nu_\rho'\,,\\
s_{\myvec\theta\cdot\myvec\nu'}(\myvec r,\myvec r')&=
-\nu_\rho'\sin(\theta-\theta')\,,\\
s_{\myvec\theta\cdot\myvec\theta'}(\myvec r,\myvec r')&=
\cos(\theta-\theta')\,,\\
s_{\myvec\theta\cdot\myvec\tau'}(\myvec r,\myvec r')&=
-\nu_z'\sin(\theta-\theta')\,.
\end{align}

Here follow $d_\alpha(\myvec r,\myvec r')$ for the 15 operators
$K_k^\alpha$ of $E_k$
\begin{align}
d_{\myvec\nu}(\myvec r,\myvec r')&=
\nu\cdot(r-r')+\nu_{\rho}\rho'(1-\cos(\theta-\theta'))\,,\\
d_{\myvec\nu'}(\myvec r,\myvec r')&=
\nu'\cdot(r-r')-\nu'_{\rho}\rho(1-\cos(\theta-\theta'))\,,\\
d_{\myvec\tau}(\myvec r,\myvec r')&=
\tau\cdot(r-r')+\nu_z\rho'(1-\cos(\theta-\theta'))\,,\\
d_{\myvec\tau'}(\myvec r,\myvec r')&=
\tau'\cdot(r-r')-\nu_z'\rho(1-\cos(\theta-\theta'))\,,\\
d_{\myvec\theta}(\myvec r,\myvec r')&=
\rho'\sin(\theta-\theta')\,,\\
d_{\myvec\theta'}(\myvec r,\myvec r')&=
\rho\sin(\theta-\theta')\,,\\
d_{\myvec\nu\times\myvec\nu'}(\myvec r,\myvec r')&=
(\nu_\rho'\tau\cdot r-\nu_\rho\tau'\cdot r')\sin(\theta-\theta')\,,\\
d_{\myvec\nu\times\myvec\theta'}(\myvec r,\myvec r')&=
-\tau\cdot(r-r')+(\tau\cdot r+\nu_\rho z')(1-\cos(\theta-\theta'))\,,\\
d_{\myvec\nu\times\myvec\tau'}(\myvec r,\myvec r')&=
(\nu_\rho\nu'\cdot r'+\nu_z'\tau\cdot r)\sin(\theta-\theta')\,,\\
d_{\myvec\tau\times\myvec\nu'}(\myvec r,\myvec r')&=
-(\nu_\rho'\nu\cdot r+\nu_z\tau'\cdot r')\sin(\theta-\theta')\,,\\
d_{\myvec\tau\times\myvec\tau'}(\myvec r,\myvec r')&=
-(\nu_z'\nu\cdot r-\nu_z\nu'\cdot r')\sin(\theta-\theta')\,,\\
d_{\myvec\tau\times\myvec\theta'}(\myvec r,\myvec r')&=
\nu\cdot(r-r')-(\nu\cdot r-\nu_zz')(1-\cos(\theta-\theta'))\,,\\
d_{\myvec\theta\times\myvec\nu'}(\myvec r,\myvec r')&=
\tau'\cdot(r-r')+(\tau'\cdot r'+\nu_\rho'z)(1-\cos(\theta-\theta'))\,,\\
d_{\myvec\theta\times\myvec\tau'}(\myvec r,\myvec r')&=
-\nu'\cdot(r-r')-(\nu'\cdot r'-\nu_z'z)(1-\cos(\theta-\theta'))\,,\\
d_{\myvec\theta\times\myvec\theta'}(\myvec r,\myvec r')&=
-(z-z')\sin(\theta-\theta')\,.
\end{align}

\section{Layer potentials for field evaluations}
\label{app:B}

The expressions for
field-evaluation~(\ref{eq:Efieldminus}--\ref{eq:Hfieldplus}) involve
layer potentials $\tilde S_k^\alpha$ and $\tilde K_k^\alpha$ which are
defined analogously to the operators $S_k^\alpha$ and $K_k^\alpha$ in
Appendix~\ref{app:A}. The only difference is that $\tilde S_k^\alpha
g(\myvec r)$ and $\tilde K_k^\alpha g(\myvec r)$ have $\myvec
r\in\Omega_-\cup\Omega_+$ while $S_k^\alpha g(\myvec r)$ and
$K_k^\alpha g(\myvec r)$ have $\myvec r\in\Gamma$. Therefore $\tilde
K_k^\alpha$ does not need the principal value.

When coding, it is helpful to have $\tilde S_k^\alpha$ and $\tilde
K_k^\alpha$ in the form
\begin{align}
\tilde S_k^\alpha g({\myvec r})&=
{\rm i}k\int_\Gamma \frac{s_\alpha(\myvec r,\myvec r')}
                 {2\pi\vert\myvec r-\myvec r'\vert}
e^{{\rm i}k\vert\myvec r-\myvec r'\vert}
g({\myvec r'})\,{\rm d}\Gamma'\,,
\quad \myvec r\in\Omega^-\cup\Omega^+\,,
\\
\tilde K_k^\alpha g({\myvec r})&=
\int_\Gamma \frac{d_\alpha(\myvec r,\myvec r')}
                 {2\pi\vert\myvec r-\myvec r'\vert^3}
(1-{\rm i}k\vert\myvec r-\myvec r'\vert)
e^{{\rm i}k\vert\myvec r-\myvec r'\vert}
g({\myvec r'})\,{\rm d}\Gamma'\,,
\quad \myvec r\in\Omega^+\cup\Omega^-\,,
\end{align}
where $s_\alpha(\myvec r,\myvec r')$ and $d_\alpha(\myvec r,\myvec
r')$ are static kernel factors, some of which are already listed
Appendix~\ref{app:A}. The static kernel factors needed, not listed in
Appendix~\ref{app:A}, are
\begin{align}
s_{\myvec\rho\cdot\myvec\nu'}(\myvec r,\myvec r')&=
 \nu_\rho'\cos(\theta-\theta')\,,\\
s_{\myvec\rho\cdot\myvec\tau'}(\myvec r,\myvec r')&=
 \nu_z'\cos(\theta-\theta')\,,\\
s_{\myvec\rho\cdot\myvec\theta'}(\myvec r,\myvec r')&=
 \sin(\theta-\theta')\,,\\
s_{\myvec z\cdot\myvec\nu'}(\myvec r,\myvec r')&=
 \nu_z'\,,\\
s_{\myvec z\cdot\myvec\tau'}(\myvec r,\myvec r')&=
-\nu_\rho'\,,
\end{align}
and
\begin{align}
d_{\myvec\rho}(\myvec r,\myvec r')&=
 \rho-\rho'\cos(\theta-\theta')\,,\\
d_{\myvec z}(\myvec r,\myvec r')&=
 z-z'\,,\\
d_{\myvec\rho\times\myvec\nu'}(\myvec r,\myvec r')&=
-(\tau'\cdot r'+\nu_\rho'z)\sin(\theta-\theta')\,,\\
d_{\myvec\rho\times\myvec\tau'}(\myvec r,\myvec r')&=
 (\nu'\cdot r'-\nu_z'z)\sin(\theta-\theta')\,,\\
d_{\myvec\rho\times\myvec\theta'}(\myvec r,\myvec r')&=
 (z-z')\cos(\theta-\theta')\,,\\
d_{\myvec z\times\myvec\nu'}(\myvec r,\myvec r')&=
 \nu_\rho'\rho\sin(\theta-\theta')\,,\\
d_{\myvec z\times\myvec\tau'}(\myvec r,\myvec r')&=
 \nu_z'\rho\sin(\theta-\theta')\,,\\
d_{\myvec z\times\myvec\theta'}(\myvec r,\myvec r')&=
 \rho'-\rho\cos(\theta-\theta')\,.
\end{align}

\begin{small}
\bibliographystyle{abbrv}
\bibliography{HKR}
\end{small}

\end{document}